\documentclass[12pt,a4paper]{JHEP3}
\usepackage[fleqn]{amsmath}
\usepackage{hhline}
\usepackage[T1]{fontenc}
\usepackage{bbm} 
\usepackage{fleqn}
\usepackage{mathrsfs} 
\usepackage{cite} 
\usepackage{longtable}
\usepackage{caption}
\usepackage{axodraw}

\DeclareMathOperator{\tr}{tr}
\DeclareMathOperator{\Tr}{Tr}
\DeclareMathOperator{\diag}{diag}

\renewcommand{\d}{\text{d}}
\newcommand{\I}{\text{i}}

\newcommand{\E}[1]{\ensuremath{\text{E}_{#1}}} 
\renewcommand{\G}[1]{\ensuremath{\text{G}_{#1}}}
\newcommand{\SO}[1]{\ensuremath{\text{SO}\!\left(#1\right)}}
\newcommand{\SU}[1]{\ensuremath{\text{SU}\!\left(#1\right)}}
\newcommand{\U}[1]{\ensuremath{\text{U}\!\left(#1\right)}}
\newcommand{\Z}[1]{\ensuremath{\mathbbm{Z}_{#1}}} 

\newcommand{\T}{\ensuremath{\boldsymbol{10}}}
\newcommand{\F}{\ensuremath{\boldsymbol{5}}}
\newcommand{\Tb}{\ensuremath{\boldsymbol{\overline{10}}}}
\newcommand{\Fb}{\ensuremath{\boldsymbol{\bar{5}}}}

\newcommand{\Xb}{\ensuremath{\bar{X}}}
\newcommand{\Yb}{\ensuremath{\bar{Y}}}
\newcommand{\Zb}{\ensuremath{\bar{Z}}}
\newcommand{\jvs}{\rule[-7pt]{0.00pt}{20pt}}

\numberwithin{equation}{section}
\numberwithin{table}{section}
\allowdisplaybreaks

\author{W.~Buchm\"uller$^a$, C.~L\"udeling$^b$, J.~Schmidt$^a$\\a Deutsches
  Elektronen-Synchrotron DESY, Hamburg, Germany\\ 
b Institut f\"ur Theoretische Physik, Universit\"at Heidelberg, Heidelberg, Germany}

\abstract{We construct a 6D supergravity theory which emerges as intermediate step
in the compactification of the heterotic string to the supersymmetric
standard model in four dimensions. The theory has $\mathcal{N}=2$ 
supersymmetry and a gravitational sector with one tensor and two 
hypermultiplets in addition to the supergravity multiplet. Compactification 
to four dimensions occurs on a $T^2/\Z2$ orbifold which has two 
inequivalent pairs of fixed points with unbroken $\SU5$ and $\SU2\times\SU4$ 
symmetry, respectively. All gauge, gravitational and mixed anomalies
are cancelled by the Green-Schwarz mechanism. The model has partial 6D 
gauge-Higgs 
unification. Two quark-lepton generations are localized at the $\SU5$ branes,
the third family is composed of split bulk hypermultiplets. The top Yukawa
coupling is given by the 6D gauge coupling, all other Yukawa couplings
are generated by higher-dimensional operators at the $\SU5$ branes. The
presence of the $\SU2\times\SU4$ brane breaks $\SU5$ and generates
split gauge and Higgs multiplets with $\mathcal{N}=1$ supersymmetry in 
four dimensions.
The third generation is obtained from two split $\boldsymbol{\bar{5}}$-plets 
and two split $\boldsymbol{10}$-plets, which together have the quantum
numbers of one $\boldsymbol{\bar{5}}$-plet and one $\boldsymbol{10}$-plet.
This avoids unsuccessful $\SU5$ predictions for Yukawa couplings of 
ordinary 4D $\SU5$ grand unified theories.}

\title{Local SU(5) Unification from the Heterotic String }

\keywords{Superstrings and Heterotic Strings, Superstring Vacua}
\preprint{\arXivid{0707.1651}\\DESY-07-072\\ HD-THEP-07-14\\JHEP {\bfseries 0709}(2007)113}

\begin{document}

\section{Introduction}\label{sec:Intro}

The symmetries and the particle content of the standard model point towards 
grand unified theories (GUTs). The simplest unified gauge group is $\SU5$ with
three $\Fb$- and $\T$-plets for the three quark-lepton generations of
the standard model \cite{gg74}. Higgs doublets can be obtained from further
$\F$- and $\Fb$-plets, with their heavy color triplet partners decoupled 
from the low energy theory. In supersymmetric GUTs the hierarchy between the 
electroweak scale and the GUT scale is stabilized and, for the minimal case 
of two Higgs doublets, gauge couplings unify at the scale
$M_\mathrm{GUT} \simeq 2\times 10^{16}\ \mathrm{GeV}$. 

Neutrino masses and mixings can be described by adding a non-renormalizable,
lepton-number violating dimension-5 operator composed of lepton and Higgs 
doublets, with coupling strength $1/\Lambda$. The observed smallness of 
the neutrino masses then requires $\Lambda = {\cal O}(M_\mathrm{GUT})$,
hinting at a $B-L$ breaking scale of the order of $M_\mathrm{GUT}$.  
Embedding $\SU5$ and $\U1_{B-L}$ in $\SO{10}$ \cite{geo75,fm75},
and continuing the route of unification via exceptional groups, one arrives
at $\E8$, which is beautifully realized in the heterotic string 
\cite{ghx851,ghx852}.

An elegant scheme leading to chiral gauge theories in four dimensions is 
the compactification on orbifolds \cite{dhx85,dhx86,inq87,ikx87,bl99}.
Recently, considerable progress has been made in deriving unified field
theories from orbifold compactifications of the heterotic string 
\cite{krz04,fnx04,krz05,ht04,bhx04,bhx052}, and it has been demonstrated 
that the idea 
of local grand unification can serve as a guide to find string vacua 
corresponding to the supersymmetric standard model \cite{bhx05,bhx06,lnx06}.
In this paper we study in some detail an orbifold GUT limit of the model 
\cite{bhx05}, where two of the compact dimensions are larger than 
the other four. In this way we hope to obtain a better understanding of
some open questions of current orbifold compactifications: the large vacuum
degeneracy, the decoupling of unwanted massless states and the 
stabilization of moduli fields.

The model \cite{bhx05} is based on a $\Z{6-\mathrm{II}}$ twist which is the 
product of a $\Z{3}$ twist and a $\Z{2}$ twist. In a first step, described
in Section~2, we 
compactify the $\E8\times\E8$ heterotic string on the orbifold 
$T^4/\Z3$, where $T^4$ is a 4-torus with the Lie algebra 
lattice $\G2 \times \SU3$. The six-dimensional (6D) theory has 
$\mathcal{N}=2$ supersymmetry and unbroken gauge group  
\begin{equation}
{\rm G_6}~=~\SU6\times\U1^3\times \left[\SU3\times\SO8\times\U1^2\right]\;,
\end{equation}
where the brackets denote the subgroup of the second $\E8$. The gravitational
sector contains one tensor multiplet whose (anti-)self-dual part belongs to
the $\mathcal{N}=2$ (dilaton) supergravity multiplet.

Compactification from six to four dimensions on the orbifold 
$T^2/\Z2$ with $\SO4$ Lie lattice leads to additional fixed points 
and twisted sectors. The massless spectrum in four dimensions agrees with the 
results obtained in \cite{bhx05,bhx06}. In addition to the zero modes, the 6D 
field theory contains the Kaluza--Klein excitations of the large $\SO4$-plane
and further non-Abelian singlets. As
described in Section~3, the projection conditions for physical massless states 
of the model \cite{bhx05} now become $\Z2$ projection conditions for 
the 6D bulk fields at the orbifold fixed points in the $\SO4$--plane. 

Given the $\Z2$ parities of the 6D bulk fields, one can perform a highly 
non-trivial consistency check of the 6D field theory, the cancellation of 
all gauge, gravitational and mixed anomalies by the Green-Schwarz mechanism 
\cite{gs84}. In Section~4 it is explicitly shown that all irreducible anomalies
vanish and that the reducible ones are indeed cancelled by a unique
Green--Schwarz term in the effective action \cite{erl94,lnz04}. The 6D
theory has different local anomalous $\U1$ symmetries at the different
fixed points in the $\SO4$ plane. Their sum yields the anomalous $\U1$ 
of the 4D theory \cite{bhx06}.

The 6D theory has a GUT gauge group and $\mathcal{N}=2$ supersymmetry, and
therefore considerably fewer multiplets than the 4D theory. This simplifies
the decoupling of unwanted exotic states as we show in Section~5. For a vacuum
with spontaneously broken $B-L$ symmetry we then obtain a local $\SU5$ GUT
model with two localized and two bulk quark-lepton families. The Higgs
fields are identified as bulk fields with partial gauge-Higgs unification.
The $\SU5$ invariant Yukawa couplings and the $\SU5$ breaking by the $\Z2$
orbifolding are discussed in Section~6. Open problems concerning supersymmetric
vacua and the stabilization of the compact dimensions are outlined
in Section~7.

Finally, in Section~\ref{sec:Outlook}, we conclude with a brief outlook on open
questions and further challenges for realistic compactifications of the
heterotic string.

\section{6D Supergravity from the Heterotic String}
\label{sec:6dsugra}

\subsection[The Heterotic String on $T^6/\Z{6-\mathrm{II}}$]{The Heterotic String on
  \boldmath$T^6/\Z{6-\mathrm{II}}$} 
We consider the propagation of the $\E8\times\E8 $  heterotic  string in a
space-time background which is the product of four-dimensional Minkowski space 
and a six-dimensional orbifold \cite{ft04}. The compact space is obtained by 
dividing
the torus $T^6 = \mathbbm{R}^6/2\pi\Lambda$ by the discrete
symmetry $\Z{6-\mathrm{II}}=\Z3\times\Z2$ of the Lie algebra lattice
 $\SO4 \times \SU3 \times \G2$. 
The four complex coordinates $z^i$, $i=1\ldots 4$, comprise the two transverse
dimensions of Minkowski space ($i=4$) and the six compact dimensions 
($i=1\ldots 3$).

The $\Z{6-\mathrm{II}}$ orbifold with the $\G2 \times \SU3 \times \SO4$ lattice
is characterized by the twist vector
\begin{equation}\label{eq:v6}
 v_6 = \left(-\frac{1}{6},-\frac{1}{3},\frac{1}{2};0\right)\;,
\end{equation}
which is the sum of $\Z3$ and $\Z2$ twist vectors, $v_6 = - v_3 + v_2$, where
\begin{equation}
v_3 = 2 v_6\;, \quad v_2 = 3 v_6\;.
\end{equation}
Note that the $\Z3$ twist leaves the $\SO4$ plane invariant whereas the
$\Z2$ twist does not affect the $\SU3$ plane. Both twists act non-trivially
on the $\G2$ plane.

In the light-cone gauge the heterotic string can be described by 4 complex  
coordinates $Z^i(\sigma)$ ($i=1\ldots4$), 4 bosonized right-moving 
Neveu-Schwarz-Ramond (NSR) fermions $H^i(\sigma_-)$ ($i=1\ldots 4$) and 16 
left-moving bosons  $X^I(\sigma_+)$ ($I=1\ldots 16$), where 
$\sigma_{\pm} = \tau \pm \sigma$. 
The fields $X^I$ are compactified  on the 16--dimensional $\E8\times\E8$ 
torus. Correspondingly, the momenta of the right-moving fields
$H^i$ lie on the weight lattice of the little group $\SO8$. 
The quantum numbers of a string state
are thus given by the $\E8\times\E8$ root vector $p^I$ for the gauge and the $\SO8$ weight vector
$q^i$ for the Lorentz quantum numbers.

The orbifold twist is embedded into the gauge group by the $\Z6$ twist vector
\begin{align}\label{eq:V6}
  V_6&=\left(-\frac{1}{2},-\frac{1}{2},\frac{1}{3},0^5\right)
  \left(\frac{17}{6},\left(-\frac{5}{2}\right)^6,\frac{5}{2},\right)\,.  
\end{align}
In addition, there are two Wilson lines associated with the two subtwists: 
a $\Z3$ Wilson line $W_3$ in the $\SU3$ plane and a $\Z2$ Wilson line $W_2$ 
in the $\SO4$ plane, given by 
\begin{align}\label{eq:W3}
  W_3&=\left(-\frac{1}{6},\frac{1}{2},\frac{1}{2},
    \left(-\frac{1}{6}\right)^5\right) \left(0,
    -\frac{2}{3},\frac{1}{3},\frac{4}{3},-1,0^3\right)\,,\\
  W_2&=\left(-\frac{1}{2},0,-\frac{1}{2},\frac{1}{2},\frac{1}{2},0^3\right)
  \left(\frac{23}{4},-\frac{25}{4}, -\frac{21}{4},-\frac{19}{4},
    -\frac{25}{4},-\frac{21}{4},
    -\frac{17}{4},\frac{17}{4}\right)\,. \label{eq:W2} 
\end{align}

A basis in the Hilbert space of the quantized string is obtained by acting 
with the creation operators ($n < 0$) for right-handed modes ($\alpha^i_{n}, 
\widetilde{\beta}^i_{n}$) and left-handed modes ($\widetilde{\alpha}^i_{n},
\widetilde{\alpha}^I_{n}$) on the ground states of the untwisted 
sector $U$ ($k=0$) and the twisted  sectors $T_k$ ($k=1\ldots 5$). 
The ground states of the different sectors depend on the momentum
vectors $q^i$, $p^I$ and, for the twisted sectors, also on the fixed point $f$
(cf.~\cite{ft04,bhx06}),
\begin{equation}
 |q,p\rangle~\equiv~|q\rangle \otimes |p\rangle\;, \quad
 |f;q,p\rangle~\equiv~|q_{\rm sh}\rangle \otimes |p_{\rm sh}\rangle\;,
\end{equation}
with the shifted momenta
\begin{equation}
q_{\rm sh} = q + kv_6\;, \quad p_{\rm sh} = p +V_f\;.
\end{equation}
Here $k$ is the order of the twist and $V_f$ is the local gauge twist at
the fixed point $f$. It turns out that for the considered model 
only oscillator modes of the left-moving strings $Z^i_\mathrm{L}(\sigma_+)$, 
$Z^{*i}_\mathrm{L}(\sigma_+)$ and $X^I(\sigma_+)$ are relevant. 

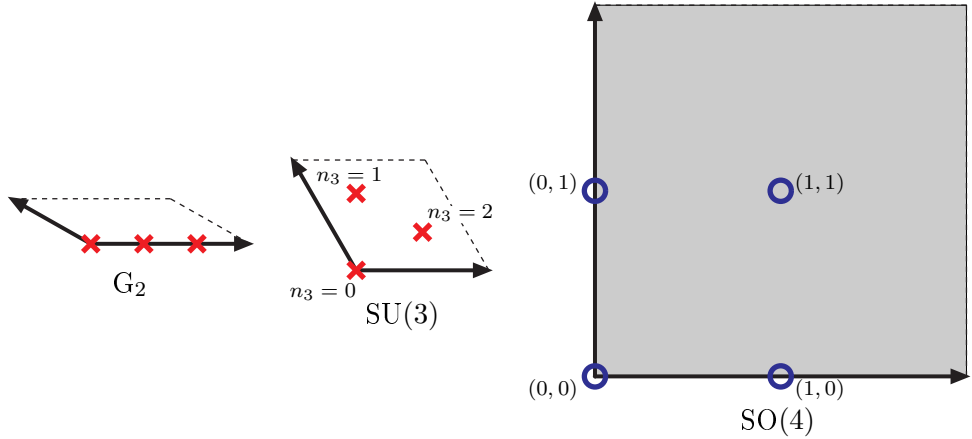
\begin{figure}[htbp]
  \begin{center}
    \begin{picture}(400,160) \small    
      {\SetWidth{1.5}
      \LongArrow(50,60)(110,60)
      \LongArrow(50,60)(20,77)}
      \DashLine(110,60)(80,77){2}
      \DashLine(20,77)(80,77){2}
      \Vertex(50,60){1}
      {\SetWidth{2}\SetColor{Red}
        \Line(47,57)(53,63)
        \Line(47,63)(53,57)
        \Line(67,57)(73,63)
        \Line(67,63)(73,57)
        \Line(87,57)(93,63)
        \Line(87,63)(93,57)
      }
      \Text(65,45)[]{$\G2$}
      {\SetWidth{1.5}
      \LongArrow(150,50)(126,91.6)
      \LongArrow(150,50)(200,50)}
      \DashLine(200,50)(176,91.6){2}
      \DashLine(126,91.6)(176,91.6){2}
      \Vertex(150,50){1}
      \Text(138,42)[]{{\scriptsize $n_3=0$}}
      \Text(148,86)[]{{\scriptsize $n_3=1$}}
      \CBox(183,68)(192,74){White}{White}
      \Text(190,72)[]{{\scriptsize $n_3=2$}}
      {\SetColor{Red}\SetWidth{2}
        \Line(147,76)(153,82)
        \Line(147,82)(153,76)
        \Line(147,47)(153,53)
        \Line(147,53)(153,47)
        \Line(172,61.4)(178,67.4)
        \Line(172,67.4)(178,61.4)
      }
      \Text(168,33)[]{$\SU3$}
      {\SetWidth{0}\GBox(240,10)(380,150){.8}}
      \DashLine(380,150)(380,10){2}
      \DashLine(380,150)(240,150){2}
      {\SetWidth{1.5}
      \LongArrow(240,10)(380,10)
      \LongArrow(240,10)(240,150)}
      \Vertex(240,10){.8}
      {\SetColor{Blue}\SetWidth{2}
        \CArc(240,10)(4,0,360)
        \CArc(240,80)(4,0,360)
        \CArc(310,10)(4,0,360)
        \CArc(310,80)(4,0,360)
      }
      \Text(225,5)[]{\scriptsize $(0,0)$}
      \Text(225,83)[]{\scriptsize $(0,1)$}
      \Text(326,5)[]{\scriptsize $(1,0)$}
      \Text(326,83)[]{\scriptsize $(1,1)$}
      \Text(310,-7)[]{$\SO4$}
    \end{picture}
  \end{center}
  
  \caption{The tori of the orbifold $T^6/\Z6$. Red crosses mark fixed points of
    the $\Z3$ twist used for the first step of  compactification. The $\SO4$
    torus is invariant, while the other tori contain three fixed points
    each. The fixed points in the $\G2$ torus are equivalent, while the $\SU3$
    torus contains a Wilson line, and the fixed points are inequivalent and
    labelled by $n_3$. The blue circles mark the $\Z2$ fixed points in the
    $\SO4$ plane which are labelled by $(n_2,n_2')$. There are further $\Z2$
    fixed points in the $\G2$ torus which are not shown. \label{fig:torusgeometry}}
\end{figure}

\subsection[Intermediate $\Z3$ Compactification]{Intermediate \boldmath$\Z3$ Compactification}
We are now interested in the effective field theory for the massless states
in the limit where the $\SO4$ plane is much larger the $\G2$ and $\SU3$ planes,
yielding approximately flat 6D Minkowski space. Hence, in a first step, we
consider the compactification on the orbifold $T^4/\Z3$. The physical
states of the gravitational sector,
\begin{equation}\label{grav}
|q,i\rangle = |q\rangle \otimes \widetilde{\alpha}^i_{-1}|0\rangle \;, \quad
|q,i^*\rangle = |q\rangle \otimes \widetilde{\alpha}^{*i}_{-1}|0\rangle \;,
\end{equation}
have to satisfy the mass equations
\begin{subequations}
\begin{eqnarray}
\frac{1}{8}m_\mathrm{R}^2  &=& \frac{1}{2} q^2 -\frac{1}{2} = 0\;,\\
\frac{1}{8}m_\mathrm{L}^2  &=& \frac{1}{2} p^2 -1 
+ \widetilde{N}+\widetilde{N}^* = 0\;.
\end{eqnarray}
\end{subequations}
Here $p = 0$, and $\widetilde{N},\widetilde{N}^*$ are the oscillator numbers
for left-moving modes in $z^i$, $z^{*i}$ directions, summed over $i$: 
$\widetilde{N} = \sum_i \widetilde{N}_i$, 
$\widetilde{N}^* = \sum_i \widetilde{N}^*_i$.
Furthermore, physical states have to be invariant under the $\Z3$ twist, 
\begin{equation}  \label{phase}
    v_3\cdot\left(\widetilde{N}-\widetilde{N}^* - q \right)~=~0\mod 1\;.
\end{equation}

The 16 bosonic states\footnote{Underline denotes
all permutations.} $q=(0,0,{\underline {\pm 1,0}})$ with $i=3,4$, together with the 16 fermionic states 
$q=\left(\frac{1}{2},\frac{1}{2},\pm\frac{1}{2},\pm\frac{1}{2}\right),
\left(-\frac{1}{2},-\frac{1}{2},\pm\frac{1}{2},\pm\frac{1}{2}\right)$ with $i=3,4$,
form the familiar 6D supergravity and dilaton $N=2$ multiplets \cite{ns86},
\begin{equation}
(G_{MN}, B_{MN}^+, \Psi_M)\; , \quad (B_{MN}^-, \Phi, \chi)\; .
\end{equation} 
Here $B_{MN}^+$ ($B_{MN}^-$) is the antisymmetric tensor field with
(anti-)self-dual field strength. Note that together there is only one tensor
field $B_{MN}$ without self-duality conditions, which is the special case for
which a lagrangian exists.

The 4 bosonic states $q=(1,0,0,0),(0,-1,0,0)$ with 
$\widetilde{N}_1 =1, \widetilde{N}_2^* = 0$ or
$\widetilde{N}_1 =0$, \mbox{$\widetilde{N}_2^* = 1$}, 
together with the corresponding 4 fermionic
states  $q=\left(\frac{1}{2},-\frac{1}{2},
  {\underline{\frac{1}{2},-\frac{1}{2}}}\right)$ and the charge conjugate
states correspond to two 6D hypermultiplets, 
\begin{equation}
C_1\;, \quad C_2\;.
\end{equation}
They contain the two `radion' fields of the small $\G2$ and $\SU3$ tori as
well as off-diagonal components of the metric and the tensor fields and the
associated superpartners. The complex structure of the small dimensions is 
fixed. All 24 bosonic
fields originate from the 64 bosonic states $\hat{G}_{MN}$, $\hat{B}_{MN}$ 
and $\hat{\Phi}$ in 10 dimensions. The remaining 40 bosonic states and their
fermionic superpartners are projected out by the $\Z3$ twist. 

The massless physical states of the gauge sector, 
\begin{equation}
|q,p\rangle~\equiv~|q\rangle \otimes |p\rangle \;,
\end{equation}
have vanishing oscillator numbers and satisfy the projection conditions 
\begin{equation}\label{proj1}
 v_3\cdot q - V_f\cdot p~=~0\mod 1\;. 
\end{equation}
Here $V_f=2 (V_6 + n_3 W_3)$ are the local $\Z3$ gauge subtwists of the model.
They differ by multiples of the $\Z3$ Wilson line $W_3$ in the $\SU3$ plane, 
which distinguishes the three inequivalent fixed points labelled by 
$n_3=0,1,2$ (cf.~Fig.~1).
Eqs.~(\ref{proj1}) are equivalent to 
\begin{equation}\label{proj2}
v_3\cdot q - V_3\cdot p~=~0~{\rm mod}~1\;, \quad W_3\cdot p = 0~{\rm mod}~1\;,
\end{equation}
where the second condition reflects the fact that the finite extension of
the $\SU3$ plane is neglected in the 6D effective field theory.

At each fixed point in the $\SU3$ plane the group $\E8\times\E8$ is 
broken to the subgroup $\SO{14}\times \U1\times\left[\SO{14}\times\U1\right]$,
which is differently embedded into $\E8\times\E8$ at the different fixed points
\cite{bhx06}. The brackets denote the subgroup  of the second $\E8$. 
The $\U1$ factors are sometimes omitted; they can always be reconstructed 
since the rank of the gauge group is preserved. One easily 
verifies that the intersection of the three $\E8\times\E8$ subgroups, which 
yields the unbroken gauge group of the 6D theory, is given by 
\begin{equation}\label{g6}
{\rm G_6}~=~\SU6\times\U1^3\times \left[\SU3\times\SO8\times\U1^2\right]\;,
\end{equation}
with the massless $\mathcal{N}=2$ vector multiplets
\begin{equation}
\left(\boldsymbol{35};1,1\right) + \left(1;\boldsymbol{8},1\right)
+ \left(1;1,\boldsymbol{28}\right) + 5\times \left(1;1,1\right)\;.
\end{equation}

The massless vector states are obtained from the conditions (\ref{proj1})
for $v_3\cdot q = 0$. There are two further possibilities,  
$v_3\cdot q = \pm 1/3$ and  $v_3\cdot q = \pm 2/3$, which lead to
$\mathcal{N}=2$ hypermultiplets. A straightforward calculation yields the gauge
multiplets 
\begin{equation}
\left(\boldsymbol{20};1,1\right) +
\left(1;1,\boldsymbol{8}\right) +
\left(1;1,\boldsymbol{8}_s\right) + \left(1;1,\boldsymbol{8}_c\right) +
4\times \left(1;1,1\right) \;,
\end{equation}
with the $\U1$ charges listed in Table~\ref{tab:torus_untwisted}.

In addition to the vector and hypermultiplets from the untwisted sector of
the string, there are 6D bulk fields which originate from the twisted
sectors $T_2$ and $T_4$ of the $\Z{6-\mathrm{II}}$ model, corresponding
to the twisted sectors $\hat{T}_1$ and $\hat{T}_2$ of the $\Z3$ subtwist.
The projection conditions for physical states are  
\begin{equation}\label{cond}
 v_3\cdot \left(\widetilde{N}_f - \widetilde{N}_f^*\right)
 - v_3\cdot \left(q+v_3\right) + V_f \cdot \left(p+V_f\right)~=~0\mod 1\;,
\end{equation}
where $\widetilde{N}_f,\widetilde{N}_f^*$ are the integer oscillator numbers
for left-moving modes localized at the fixed point $f$ (cf.~\cite{bhx06}).

  \begin{table}[t]
    \centering
    \begin{tabular}{|c|c|c|c|}\hhline{----}
      Sector    & Multiplet & Representation                     & \# \\ \hhline{====}
      Gravity   & Graviton  &        $G_{MN}$                    & 1 \\
                & Dilaton   &        $\Phi$                      & 1 \\
                & Hyper     &        $C_1$, $C_2$                & 2 \\\hhline{----}
      Untwisted & Vector    & $\left(\boldsymbol{35};1,1\right)$ & 35 \\
                &           & $\left(1;\boldsymbol{8},1\right)$  & 8 \\
                &           & $\left(1;1,\boldsymbol{28}\right)$ & 28 \\
                &           & $5\times \left(1;1,1\right)$       & 5 \\\hhline{----}
      Untwisted & Hyper     & $\left(\boldsymbol{20};1,1\right)$ & 20\\
                &           & $\left(1;1,\boldsymbol{8}\right) +\left(1;1,\boldsymbol{8}_s\right)
                            +\left(1;1,\boldsymbol{8}_c\right) $ & 24\\ 
                &           & $4\times \left(1;1,1\right)$ & 4\\\hhline{----}
      Twisted   & Hyper     & $9\times\left(\boldsymbol{6};1,1\right)+9\times
                               \left(\boldsymbol{\bar{6}};1,1\right)$ & 108 \\
                &           & $9\times\left(1;\boldsymbol{3};1,1\right)+9\times
                                \left(1;\boldsymbol{\bar{3}};1,1\right)$ & 54 \\
                &           & $3\times\left(1;1,\boldsymbol{8}\right)
                            +3\times\left(1;1,\boldsymbol{8}_s\right) +
                            3\times\left(1;1,\boldsymbol{8}_c\right)$ & 72\\
                &           & $36\times\left(1;1,1\right)$ & 36\\\hhline{----}
    \end{tabular}
    \caption{$\mathcal{N}=2$ supermultiplets of the 6D theory: graviton, dilaton, 76 vector
    and 320 hypermultiplets. The non-Abelian symmetry group is 
    $\SU6\times[\SU3\times\SO8]$.}
   \label{tab:torusfieldsshort}
  \end{table}

At each fixed point one has states with $\widetilde{N}_f=\widetilde{N}_f^*=0$,
which yield $\mathcal{N}=2$ hypermultiplets $(\boldsymbol{14},1)$ and
$(1,\boldsymbol{14})$. With respect to the 6D gauge group these multiplets
form the reducible representations 
\begin{subequations}
\begin{align}
(\boldsymbol{14},1) &= (\boldsymbol{6};1,1) + (\boldsymbol{\bar{6}};1,1)
+ 2\times (1;1,1)\;, \\
(1,\boldsymbol{14}) &= (1;\boldsymbol{3},1) + (1;\boldsymbol{\bar{3}},1)
+ (1;1,\boldsymbol{\hat{8}})\;.
\end{align}
\end{subequations}
At the three $\SU3$ fixed points, $(1;1,\boldsymbol{\hat{8}})$ corresponds to
$(1;1,\boldsymbol{8})$, $(1;1,\boldsymbol{8}_s)$ and $(1;1,\boldsymbol{8}_c)$,
respectively. Furthermore, there are oscillator states for the two small
compact planes,
\begin{equation}
|q+v_3\rangle \otimes \widetilde{\alpha}^i_{f-1}|p+V_f\rangle \;, \quad
|q+v_3\rangle \otimes \widetilde{\alpha}^{*i}_{f-1}|p+V_f\rangle \;, \quad
i = 3,4\;,
\end{equation}
which yield two non-Abelian singlet hypermultiplets for each fixed point.

In addition to the three inequivalent fixed points in the $\SU3$ plane,
there are three equivalent fixed points of the $\Z3$ twist in the $\G2$ plane.
This yields a multiplicity of three for all hypermultiplets from the $T_2$ and
$T_4$ sectors. All the multiplets of the 6D theory are summarized in
Table~\ref{tab:torusfieldsshort}. The full listing including the $\U1$
charges is given in Appendix~\ref{app:bulkstates}.

Let us finally consider the interaction between vector and hypermultiplets.
It is convenient to decompose all $\mathcal{N}=2$ 6D multiplets in terms of
$\mathcal{N}=1$ 4D 
multiplets. The 6D vector multiplet splits into a pair of 4D vector and chiral
multiplets, \mbox{$A = (V,\phi)$}, and a hypermultiplet consists of a pair of chiral
multiplets, $H = (H_L, H_R)$; here $\phi$ and $H_L$ are left-handed,
$H_R$ is right-handed. In flat space, the
interaction lagrangian takes the simple form \cite{ahx01} 
\begin{align}
  \begin{split}
    {\cal L}_H &= \int d^4\theta\left(H_L^\dagger e^{2gV} H_L 
      + H_R^{c\dagger} e^{-2gV} H_R^c\right)\\
    &\quad + \int d^2\theta\ H_R^c\left(\partial +\sqrt{2}g\phi\right)H_L + \text{h.c.}
  \end{split}\label{tyuk1}
\end{align}
After compactification to four dimensions, the first term yields the familiar
gauge interactions, whereas the second term can give rise to Yukawa couplings.
For the hypermultiplet $(\boldsymbol{20};1,1)$ one obtains
\begin{equation}
{\cal L}_H \supset \sqrt{2}g\int d^2\theta\ H_R^c(\boldsymbol{20})
\phi(\boldsymbol{35})H_L(\boldsymbol{20}) + \text{h.c.} 
\end{equation}
The $\SU6$ $\boldsymbol{20}$-plet contains $\SU5$ $\T$- and
$\Tb$-plets, and the $\boldsymbol{35}$-plet contains
$\SU5$ $\F$- and $\Fb$-plets. As we shall see in
Section~6, after projection onto 4D zero modes, Eq.~(\ref{tyuk1}) 
yields precisely the top Yukawa coupling.
The Yukawa terms for the  hypermultiplets $(\boldsymbol{6};1,1)$ and
$(\boldsymbol{\bar{6}};1,1)$,
\begin{equation}
{\cal L}_H \supset \sqrt{2}g\int d^2\theta \left( H_R^c(\boldsymbol{6})
\phi(\boldsymbol{35})H_L(\boldsymbol{6}) + H_R^c(\boldsymbol{\bar{6}})
\phi(\boldsymbol{35})H_L(\boldsymbol{\bar{6}})\right) + \text{h.c.} 
\end{equation}
will be important for the decoupling of exotic states in Section~5.


\section{\boldmath$\Z2$ Compactification to Four Dimensions}
\label{sec:z2comp}

The compactification from six to four dimensions on a $\Z2$ orbifold leads to
four additional fixed points in the $\SO4$ plane and to further projection
conditions for physical massless states. The fixed points are labelled by
$(n_2,n_2')=(0,0),(0,1,),(1,0),(1,1)$  (cf.~Fig.~1). Due to the Wilson line 
$W_2$, they come in two pairs of equivalent fixed points, and the projection
conditions only depend on $n_2$ and not on $n_2'$.

At the fixed points, half of the supersymmetry generators are broken and only
\mbox{$\mathcal{N}=1$} supersymmetry remains unbroken. For the gravitational and gauge
multiplets of the untwisted sector the projection conditions are \cite{bhx06}
\begin{equation}
v_2 \cdot \left(\tilde{N} - \tilde{N}^*\right)  - v_2 \cdot q + V_f \cdot p 
=  0~{\rm mod}~1\;,
\label{locproj1}
\end{equation}
where $v_2 = 3v_6$, and $V_f=3V_6 + n_2 W_2$ are the local twists at the fixed 
points $n_2=0,1$ in the $\SO4$ plane.

In this paper we consider an anisotropic orbifold where the $\SO4$ plane is 
much larger than the $\G2$ and $\SU3$ planes. The Kaluza--Klein states of
the $\SO4$ plane can be included in an effective field theory below
the string scale by considering fields in the two large compact dimensions
instead of 4D zero modes which are assumed to be constant in the compact
dimensions. For the $\Z2$ twist, one has (cf.~\cite{bhx06})
$(\theta^3,l_f)(z_f^3+z^3) = z_f^3 - z^3$,  where $(\theta^3,l_f)$ is the
space group element of the fixed point $f$ and
$z^3=y^5+ \I y^6$ is the 
complex coordinate in the $\SO4$ plane. The projection conditions 
(\ref{locproj1}) for the massless
states then become local projection conditions for fields in the compact
dimensions,
\begin{align}
  \begin{split}
    P_f: \, \,\, \phi(y_f+y) &= \eta_f(\phi) \, \phi(y_f-y) \;,\\
    \eta_f(\phi) &= \exp\!\left\{2\pi\I\left(
        v_2 \cdot (\tilde{N} - \tilde{N}^* -q) + V_f \cdot p \right)\right\}\;.
  \end{split}\label{locproj2}
\end{align}
The momenta $p$, $q$ and the oscillator number $\tilde{N} - \tilde{N}^*$ 
of the states determine the quantum numbers of the corresponding
fields $\phi$, and $\eta_f(\phi) = \pm 1$. Only fields which have positive
parity at all fixed points have zero modes.

\begin{table}[t]
\begin{center}
  \begin{tabular}{|c|c|}
    \hline
    $n_2$ & Gauge group \jvs\\
    \hline
    \hline
    0 & $\SU5 \times \U1^4 \times \left[\SU3 \times \SO8 \times \U1^2 \right]$ \jvs\\
    \hline
    1 & $\SU2 \times \SU4 \times \U1^4 \times \left[ \SU2' \times \SU4' \times
      \U1^4 \right]$ \jvs\\ 
    \hline
    \hline
    $\cap$ & $\SU3 \times \SU2 \times \U1^5 \times \left[\SU2' \times \SU4'
      \times \U1^4 \right]$ \jvs\\  
    \hline
  \end{tabular}
\caption{List of the local gauge groups and their intersection.}
\label{tab:locgroups}
\end{center}
\end{table}

As an example, consider the 6D metric
\begin{equation}
  \d s^2 = g_{MN} \d x^M \d x^N = g_{\mu\nu}\d x^\mu\d x^\nu + 
  2 g_{\mu m} \d x^\mu \d y^m + g_{mn} \d y^m \d y^n\;,
\end{equation}
where $x^{\mu}$ and $y^m$ are the coordinates of 4D Minkowski space and the
two compact dimensions, respectively. One easily obtains from Eqs.~(\ref{grav})
and (\ref{locproj2}) the projection conditions
\begin{align}
g_{\mu\nu}(x,y) = g_{\mu\nu}(x,-y)\;, \quad g_{\mu m}(x,y) 
= - g_{\mu m}(x,-y)\;, \quad g_{mn}(x,y) = g_{mn}(x,-y)\;.
\end{align}
The 4D zero mode $g_{\mu\nu}(x)$ is part of the $\mathcal{N}=1$ supergravity
multiplet ($g_{\mu\nu},\psi_{\mu}$) while the three degrees of freedom in
$g_{mn}(x)$ join with $B_{56}$ to form the moduli multiplets $T$ and $S$
containing the radion field and the complex structure of the torus.

The projection conditions for the $\mathcal{N}=2$ vector multiplets $A$ are 
most conveniently expressed in terms of the corresponding $\mathcal{N}=1$
vector ($V$) and chiral ($\phi$) multiplets, $A = (V,\phi)$, which are 
elements of the Lie algebra of the 6D bulk gauge group. The unbroken gauge 
group at the fixed point $f$ is determined by the condition
\begin{equation}
p \cdot V_f = 0~{\rm mod}~1\;.
\end{equation}
At the fixed points $n_2=0$ and $n_2=1$ in the $\SO4$ plane, the bulk gauge
group $\SU6\times[\SU3\times\SO8]$ is 
broken to subgroups containing $\SU5\times[\SU2'\times\SU4']$ and 
$\SU2\times\SU4\times[\SU2'\times\SU4']$, respectively. At the two fixed 
points the conditions for the vector and chiral multiplets are given by
\begin{equation}
P_f V(x,y_f-y) P_f = V(x,y_f+y)\;, \quad
P_f \phi(x,y_f-y) P_f = -\phi(x,y_f+y)\;,
\end{equation}
where $P_f$ is the $\Z2$ parity matrix. Again only $\mathcal{N}=1$ 
supersymmetry remains unbroken. As an example, for the $\SU6$ factor, 
one has \mbox{$P_0=\diag(1,1,1,1,1,-1)$}
at $n_2=0$, and $P_1=\diag(1,1,-1,-1,-1,-1)$ at $n_2=1$.
The decomposition of the bulk gauge fields with respect to the locally
unbroken subgroups, together with all $\U1$ charges, are listed in Tables~\ref{tab:states_A_nn0}
and \ref{tab:states_A_nn1}. For the unbroken subgroup, vectors have positive and scalars
negative parity; for the broken generators the situation is reversed.

\begin{table}[t]
\begin{center}
  \begin{tabular}{|c||c||c|c||c|}
    \hline
    Bulk & $n_2=0$ &  V & $\phi$ & $t_6^0$ \jvs \\
    \hline
    \hline
    $({\bf 35};1,1)$ & $({\bf 24};1,1)$  & $+$ & $-$ & 0 \\
               & $({\bf 5};1,1) $    & $-$ & $+$ & $-6$   \\
         & $({\bf \bar{5}};1,1) $    & $-$ & $+$ & $6$  \\
         & $(1;1,1)$  & $+$ & $-$ & 0  \\
    \hline
    $(1;{\bf 8},1)$ & $(1;{\bf 8},1)$  & $+$ & $-$ & 0  \\
    \hline
    $(1;1,{\bf 28})$ & $(1;1,{\bf 28})$  & $+$ & $-$ & 0   \\
    \hline
  \end{tabular}
\caption{Local decomposition of vector multiplets at $n_2=0$. \label{tab:states_A_nn0}} 
\vspace{1cm}
  \begin{tabular}{|c||c||c|c||c|c|c|}
    \hline
     Bulk & $n_2=1$  & V & $\phi$ & $t_6^1$ & $t_7$ & $t_8$  \jvs\\
    \hline
    \hline
    $({\bf 35};1,1)$ & $({\bf 3},1;1,1)$  & $+$ & $-$ & 0 & 0 & 0  \\
                 & $(1,{\bf 15};1,1)$  & $+$ & $-$ & 0 & 0 & 0  \\
         & $({\bf 2},{\bf 4};1,1)$  & $-$ & $+$ & 15 & 0 & 0  \\
         & $({\bf 2},{\bf \bar{4}};1,1)$  & $-$ & $+$ & $-15$ & 0 & 0  \\
         & $(1,1;1,1)$  & $+$ & $-$ & 0 & 0 & 0  \\
    \hline
    $(1;{\bf 8},1)$ & $(1,1;{\bf 3},1)$  & $+$ & $-$ & 0 & 0 & 0  \\
          & $(1,1;{\bf 2},1)$  & $-$ & $+$ & 0 & 3 & 0 \\
          & $(1,1;{\bf 2},1)$  & $-$ & $+$ & 0 & $-3$ & 0  \\
          & $(1,1;1,1)$  & $+$ & $-$ & 0 & 0 & 0  \\
    \hline
    $(1;1,{\bf 28})$ & $(1,1;1,{\bf 15})$  & $+$ & $-$ & 0 & 0 & 0 \\
           & $(1,1;1,{\bf 6})$  & $-$ & $+$ & 0 & 0 & 2  \\
           & $(1,1;1,{\bf 6})$  & $-$ & $+$ & 0 & 0 & $-2$  \\
           & $(1,1;1,1)$  & $+$ & $-$ & 0 & 0 & 0  \\
    \hline
  \end{tabular}
\end{center}
\caption{Local decomposition of vector multiplets at $n_2=1$. \label{tab:states_A_nn1}} 
\end{table}

At the fixed point $n_2=0$ the GUT group $\SU5\times\U1$ is unbroken,
and the \mbox{$\mathcal{N}=2$} vector multiplet $\bf 35$ of $\SU6$ splits
into the $\mathcal{N}=1$ vector multiplets $\bf 24 + 1$ with positive parity
and the $\mathcal{N}=1$ chiral multiplets $\F + \Fb$  with positive parity 
from the coset $\SU6/(\SU5\times\U1)$.
From Table~\ref{tab:states_A_nn1} one reads off that the projection condition at
the fixed point $n_2=1$ projects out the color triplets from both the $\F$- and the 
$\Fb$-plets. This is the well known doublet-triplet splitting of orbifold GUTs.
As we shall discuss in Section~\ref{sec:decoupling}, the remaining $\SU2$ 
doublets 
can play the role of Higgs or lepton doublets in the 4D effective theory. 

The  $\mathcal{N}=2$ hypermultiplets $H$ consist of pairs of $\mathcal{N}=1$
left- and right-chiral multiplets, $H=(H_L,H_R)$. For the projection conditions
one finds
\begin{equation}
P_f H_L(x,y_f-y) = \eta_f H_L(x,y_f+y)\;, \quad
P_f H_R(x,y_f-y) = -\eta_f H_R(x,y_f+y)\;,
\end{equation}
where $P_f$ is now a matrix in the representation of $H$, and $\eta_f$
has to be calculated using Eq.~(\ref{locproj2}). The parities for the
hypermultiplets from the untwisted sector, decomposed with respect to
the unbroken groups at the fixed points $n_2=0$ and $n_2=1$ are listed in
the Tables~\ref{tab:states_BC_nn0} and \ref{tab:states_BC_nn1}.

Zero modes with standard model quantum numbers are contained in two
$\mathcal{N}=1$ chiral multiplets which are $\SU5$ $\T$-plets,
\begin{equation}
H_L = (\T;1,1)\;, \quad H_R^c = (\Tb^c;1,1)\;.
\end{equation}              
From the Tables~~\ref{tab:states_BC_nn0} and \ref{tab:states_BC_nn1} one easily
verifies that the projection
conditions at the fixed point $n_2=1$ yield the following quark-lepton
states as 4D zero modes:
\begin{equation}\label{eq:qe10}
\T:\, (3,2)=q\;; \quad \Tb^c: \, (\bar{3},1)=u^c\;,\, (1,1)=e^c\;.
\end{equation}
Together, the zero modes have again the quantum numbers of one $\SU5$
$\T$-plet. However, as we shall see in Section~6, it is crucial for their 
Yukawa couplings that they originate from two distinct $\SU5$ $\T$-plets.

\begin{table}
\begin{center}
\small
\begin{tabular}{|c||c||c|c||c|c|c|c|c|c||c|}
\hline
Bulk & $n_2=0$  & $H_L$ & $H_R$ & $t_6^0$ & $t_1$ & $t_2$ & $t_3$ &
$t_4$ & $t_5$ &  \jvs\\
\hline
\hline
$({\bf 20};1,1)$ & $({\bf 10};1,1)$  & $+$ & $-$ & 3 & $-\frac12$ & $\frac12$ & 0 & 0 & 0 &\\
\jvs     & $({\bf \bar{10}};1,1)$  & $-$ & $+$ & -3 & $-\frac12$ & $\frac12$ & 0 & 0 & 0 &\\
\hline
\jvs   $(1;1,{\bf 8})$ & $(1;1,{\bf 8})$  & $-$ & $+$ & 0 & 0 & 0 & 0 & $-1$ & 0 &\\
\hline
\jvs   $(1;1,{\bf 8_s})$ & $(1;1,{\bf 8_s})$  & $+$ & $-$ & 0 & 0 & 0 & 0 & $\frac12$ & $\frac 32$ &\\
\hline
\jvs  $(1;1,{\bf 8_c})$ & $(1;1,{\bf 8_c})$  & $+$ & $-$ & 0 & 0 & 0 & 0 & $\frac12$ & $-\frac 32$ &\\
\hline
\jvs  $(1;1,1)$ & $(1;1,1)$  & $-$ & $+$ & 0 & $\frac12$ & $\frac12$ & 3 & 0 & 0 & $U_1$\\
\hline 
\jvs   $(1;1,1)$ & $(1;1,1)$  & $+$ & $-$ & 0 & $\frac12$ & $\frac12$ & $-3$ & 0 & 0 & $U_2$\\
\hline
\jvs  $(1;1,1)$ & $(1;1,1)$  & $+$ & $-$ & 0 & $1$ & $-1$ & 0 & 0 & 0 & $U_3$\\
\hline
\jvs  $(1;1,1)$ & $(1;1,1)$  & $+$ & $-$ & 0 & $-1$ & $-1$ & 0 & 0 & 0 & $U_4$\\
\hline
\end{tabular}
\caption{Local decomposition of untwisted hypermultiplets at $n_2=0$. \label{tab:states_BC_nn0}}
\vspace*{1cm}
\begin{tabular}{|c||c||c|c||c|c|c|c|c|c|c|c||c|}
\hline
Bulk & $n_2=1$  & $H_L$ & $H_R$ & $t_6^1$ & $t_7$ & $t_8$ & $t_1$ &
$t_2$ & $t_3$ & $t_4$ & $t_5$ & \jvs \\
\hline
\hline
\jvs     $({\bf 20};1,1)$ & $({\bf 2},{\bf 6};1,1)$  & $-$ & $+$ & 0 & 0 & 0 & $-\frac12$ & $\frac12$ & 0 & 0 & 0 & \\
\jvs              & $(1,{\bf 4};1,1)$        & $+$ & $-$ & $-15$ & 0 & 0 & $-\frac12$ & $\frac12$ & 0 & 0 & 0 & \\
\jvs              & $(1,{\bf \bar{4}};1,1)$  & $+$ & $-$ & $15$ & 0 & 0 & $-\frac12$ & $\frac12$ & 0 & 0 & 0 & \\
\hline
\jvs     $(1;1,{\bf 8})$ & $(1,1;1,{\bf 4})$  & $-$ & $+$ & 0 & 0 & $-1$ & 0 & 0 & 0 & $-1$ & 0 & \\
\jvs           & $(1,1;1,{\bf \bar{4}})$  & $+$ & $-$ & 0 & 0 & $1$ & 0 & 0 & 0 & $-1$ & 0 & \\
\hline
\jvs     $(1;1,{\bf 8_c})$ & $(1,1;1,{\bf 6})$  & $-$ & $+$ & 0 & 0 & 0 & 0 & 0 & 0 & $\frac12$ & $-\frac32$ & \\
\jvs           & $(1,1;1,1)$  & $+$ & $-$ & 0 & 0 & 2 & 0 & 0 & 0 & $\frac12$ & $-\frac32$ & \\
\jvs           & $(1,1;1,1)$  & $+$ & $-$ & 0 & 0 & $-2$ & 0 & 0 & 0 & $\frac12$ & $-\frac32$ & \\
\hline
\jvs     $(1;1,{\bf 8_s})$ & $(1,1;1,{\bf 4})$  & $-$ & $+$ & 0 & 0 & $1$ & 0 & 0 & 0 & $\frac12$ & $\frac32$ & \\
\jvs           & $(1,1;1,{\bf \bar{4}})$  & $+$ & $-$ & 0 & 0 & $-1$ & 0 & 0 & 0 & $\frac12$ & $\frac32$ & \\
\hline
\jvs     $(1;1,1)$ & $(1,1;1,1)$  & $-$ & $+$ & 0 & 0 & 0 & $\frac12$ & $\frac12$ & 3 & 0 & 0 & $U_1$\\
\hline
\jvs     $(1;1,1)$ & $(1,1;1,1)$  & $-$ & $+$ & 0 & 0 & 0 & $\frac12$ & $\frac12$ & $-3$ & 0 & 0 & $U_2$\\
\hline
\jvs     $(1;1,1)$ & $(1,1;1,1)$  & $-$ & $+$ & 0 & 0 & 0 & 1 & $-1$ & 0 & 0 & 0 & $U_3$\\
\hline
\jvs     $(1;1,1)$ & $(1,1;1,1)$  & $-$ & $+$ & 0 & 0 & 0 & $-1$ & $-1$ & 0 & 0 & 0 & $U_4$\\
\hline
\end{tabular}
\end{center}
\caption{Local decomposition of untwisted hypermultiplets at $n_2=1$.\label{tab:states_BC_nn1}}
\end{table}

As discussed in the previous section, the $\mathcal{N}=2$ hypermultiplets 
from the $T_2/T_4$ sector are bulk fields in the $\SO4$ plane, but localized 
in the $\G2$ and $\SU3$ planes. With respect to the bulk gauge group they
transform as $(\boldsymbol{6};1,1)$, $(\boldsymbol{\bar{6}};1,1)$,
$(1;\boldsymbol{3},1)$, $(1;\boldsymbol{\bar{3}},1)$, 
$(1;1,\boldsymbol{8})$, $(1;1,\boldsymbol{8}_c)$,
$(1;1,\boldsymbol{8}_s)$ and $(1;1,1)$.
One can form linear combinations of the states localized at the equivalent 
fixed points in the $\G2$ plane, which are eigenstates of the $\Z2$ twist,
$\Theta^3|q_\gamma\rangle = \exp{(2\pi\I q_\gamma)}|q_\gamma\rangle$. 
For the twisted sector fields the projection conditions depend on the phase
$q_\gamma$, and the parities $\eta_f(\phi)$ are given by
\begin{equation}
\eta_f(\phi) = \exp\!\left\{2\pi\I\left(v_2 \cdot (\tilde{N} - \tilde{N}^* -q)
+ V_f \cdot p + q_\gamma \right)\right\}\;.
\end{equation}
For the $T_2/T_4$ twisted states $q_\gamma$ takes the values $0,1/2,1$. The
corresponding 6 parities for all hypermultiplets $H=(H_L,H_R)$ at the fixed
points $n_2=0$ and $n_2=1$ are listed in
Tables~\ref{tab:states_T2T4_nn0}--\ref{tab:states_T2T4star_nn1}. 

The 6D theory contains 9 hypermultiplets of $\SU5$ $\bf 5$-plets and 9
hypermultiplets of $\bf \bar{5}$-plets. Each hypermultiplet contains a 
pair of $\bf 5$ and $\bf \bar{5}$ $\mathcal{N}=1$ chiral multiplets. As
Table A.4 shows, the positive parities select from each triplet of 
hypermultiplets, with $q_\gamma = 0,1/2,1$, a chiral combination of 
$\bf 5$-plets: one $\bf 5$ and two $\bf \bar{5}$'s or two $\bf 5$'s and
one $\bf \bar{5}$. The projection conditions at $n_2=1$ then leave as 4D zero
modes from each $\bf 5$- or $\bf \bar{5}$-plet either the $\SU3$ triplet
or the $\SU2$ doublet. In this way a spectrum of massless states is generated
which is chiral with respect to the standard model group.
 
The $\Z2$ orbifolding leads from the $\Z3$ orbifold model of
Section~\ref{sec:6dsugra} to a $\Z6$ orbifold model, and therefore to new 
twisted sectors $T_1/T_5$ and $T_3$. The massless states are obtained from 
the corresponding mass equations (cf.~\cite{bhx06}) with 
$k=1$ and \mbox{$k=3$}, respectively. In the $T_3$ sector one can choose a basis of
eigenstates of the $\Z3$ twist, $\Theta^2|q_\gamma\rangle = 
\exp{(2\pi\I q_\gamma)}|q_\gamma\rangle$, with $q_\gamma = 0,1/3,
-1/3,1$ (cf.~\cite{bhx06}). The projection conditions for physical states 
now involve the parities
\begin{equation}
\eta_f(\phi) = \exp\!\left\{2\pi\I\left(v_3 \cdot (\tilde{N}-\tilde{N}^* -q)  
+ V_f \cdot p + q_\gamma \right)\right\}\;.
\end{equation}
The states are bulk fields in the $\SU3$ plane, whose extension we neglect,
but localized in the $\G2$ and $\SO4$ planes. All massless states from the
$T_1/T_5$ and $T_3$ sectors at the fixed points $n_2=0$ and $n_2=1$ are 
listed in Tables~\ref{tab:states_T1T3T5_nn0} and \ref{tab:states_T1T3T5_nn1}.

At both fixed points with $n_2=0$, one standard model family with $\SU5$ 
quantum
numbers $\bf \bar{5} + 10$ occurs. All other states are standard model 
singlets. On the contrary, there are no standard model singlets at the fixed
point $n_2=1$, but only color singlets with exotic $\SU2\times\U1$ quantum
numbers.

So far we have ignored the localization number $n_2'=0,1$ of the fixed points
in the $\SO4$ plane, since it just leads to a doubling of the states 
localized at $n_2=0,1$. Altogether, we have a rather simple picture for
the standard model non-singlet states: There are two quark-lepton
families localized at
\begin{equation}
n_2=0,\; n_2'=0,1:\quad \Fb\ +\ \T\;.
\end{equation}
From the bulk fields, vector and hypermultiplets, we have
\begin{equation}
11 \times \Fb\ + \ 9\times \F\ +\ \T\ +\ \Tb^c \;.
\end{equation} 
The spectrum is chiral and looks like four quark-lepton families plus
9 pairs of $\F$'s and $\Fb$'s. However, the projection conditions at the
$n_2=1$ fixed points eliminate half of the bulk fields, so that one is
left with three quark-lepton families and several vector-like pairs of 
$\SU3$ triplets and $\SU2$ doublets which can accommodate a
pair of Higgs doublets. Which $\Fb$'s contain the quark and lepton states
of the third family, and which one the Higgs doublet depends on the chosen
vacuum. At the fixed points $n_2=1$ there are additional localized states 
with exotic quantum numbers. Using the Tables 3.2--3.5 and A.4--A.9, one 
can check that the spectrum of zero modes obtained in \cite{bhx06} is 
reproduced.

The determination of possible supersymmetric vacua, where some of the
standard model singlet fields acquire large VEVs, is discussed in Sections~5
and 6.
In such vacua, unwanted $\SU3$ triplets and $\SU2$ can be decoupled. 
The positive and negative parities at the fixed points $n_2=0,1$, listed in 
the Tables 3.2--3.5 and A.4--A.7 are also needed to check the cancellation of 
anomalies for the constructed 6D supergravity theory.


\section{Anomalies}\label{sec:anomalies}

  \begin{table}[t]
    \centering
    \small
    \begin{tabular}{|c|l|c|c|c|}
  \hhline{-----}
  $\U1$ &Generator Embedding into $E_8 \times E_8$ & Bulk & $n_2=0$ & $n_2=1$ \jvs \\
  \hhline{=====}
  $t_1 $ & $ \left(0, 1, 0, 0, 0, 0, 0, 0\right)\left(0, 0, 0, 0, 0, 0, 0,
    0\right)$ & $\surd$ & $\surd$ & $\surd$ \jvs \\ 
  \hhline{-----}
  $t_2 $ & $ \left(0, 0, 1, 0, 0, 0, 0, 0\right)\left(0, 0, 0, 0, 0, 0, 0,
    0\right)$ & $\surd$ & $\surd$ & $\surd$ \jvs \\ 
  \hhline{-----}  
  $t_3 $ & $ \left(1, 0, 0, 1, 1, 1, 1, 1\right)\left(0, 0, 0, 0, 0, 0, 0,
    0\right)$ & $\surd$ & $\surd$ & $\surd$ \jvs \\ 
  \hhline{-----}  
  $t_4 $ & $ \left(0, 0, 0, 0, 0, 0, 0, 0\right)\left(1, 0, 0, 0, 0, 0, 0,
    0\right)$ & $\surd$ & $\surd$ & $\surd$ \jvs \\ 
  \hhline{-----}  
  $t_5 $ & $ \left(0, 0, 0, 0, 0, 0, 0, 0\right)\left(0, 1, 1, 1, 0, 0, 0,
    0\right)$ & $\surd$ & $\surd$ & $\surd$ \jvs \\ 
  \hhline{-----}  
  $t^0_6 $ & $ \left(5, 0, 0, -1, -1, -1, -1, -1\right)\left(0, 0, 0, 0, 0, 0, 0, 0\right)$ &
  $\times$ & $\surd$ & $\times$ \jvs \\  
  \hhline{-----}  
   $t^1_6 $ & $ \left(5, 0, 0, -10, -10, 5, 5, 5\right)\left(0, 0, 0, 0, 0,
    0, 0, 0\right)$ & $\times$ & $\times$ & $\surd$ \jvs \\ 
  \hhline{-----}
  $t_7 $ & $ \left(0, 0, 0, 0, 0, 0, 0, 0\right)\left(0, 1, 1, -2, 0, 0, 0,
    0\right)$ & $\times$ & $\times$ & $\surd$ \jvs \\ 
  \hhline{-----}  
  $t_8 $ & $ \left(0, 0, 0, 0, 0, 0, 0, 0\right)\left(0, 0, 0, 0, -1, -1, -1,
    1\right)$ & $\times$ & $\times$ & $\surd$ \jvs \\ 
  \hhline{-----}
  $t_{\rm an}^0 $ & $ \left(5, 0, -4, -1, -1, -1, -1, -1\right)\left(5, -1, -1, -1, 0, 0, 0,
  0\right)$ &  & $\surd$ &  \jvs \\ 
  \hhline{-----}
  $t_{\rm an}^1 $ & $ \left(1, 3, -1, 1, 1, 1, 1, 1\right)\left(-4, 4, 4, 4,
    0, 0, 0, 0\right)$ &  &  & $\surd$ \jvs \\
  \hhline{-----}
  $t_{\rm an}^{(\text{4d})} $ & $ \left(\frac{11}{6},\frac{1}{2} , -\frac{3}{2},
  -\frac{1}{6},-\frac{1}{6}, -\frac{1}{6}, -\frac{1}{6}, -\frac{1}{6}\right)\left(1,
  \frac{1}{3}, \frac{1}{3}, \frac{1}{3}, 0, 0, 0, 0\right)$ &  &  &  \jvs  \\
  \hhline{-----}
\end{tabular}
\caption{Definition of the $\U1$ generators. The last three columns
  indicate whether the generator is part of a non-Abelian group ($\times$) 
or commutes with the semi-simple group ($\surd$) in the bulk and at the fixed 
points. The anomalous $\U1$'s are linear combinations of the commuting $\U1$'s 
at the fixed point specified by the superscript
or in four dimensions; they are denoted by $t_{\rm an}^0$,
  $t_{\rm an}^1$ and $t_{\rm an}^{(\text{4d})}$, respectively.} 
\label{tab:u1defs}
\end{table}


  Anomalies of field theories on orbifolds are well understood \cite{ss04},
  and also the six-dimensional case has been discussed in
  detail \cite{erl94,abc03,gq03,lnz04,ger07}. In general the orbifold anomaly has bulk
  and brane contributions.
  While the bulk terms are already present in the torus compactification,
  the localized anomalies crucially depend on the projection conditions at
  the orbifold fixed points and the twisted sectors of the orbifold. Thus the
  requirement that all anomalies of the model can be cancelled imposes highly 
  non-trivial conditions on the spectrum. In the present model their 
  fulfillment is guaranteed by the fact that it has been
  derived from string theory, which automatically provides the right 
  Green--Schwarz terms for anomaly cancellation \cite{gs84}. In this section 
  we apply its six-dimensional version \cite{erl94,lnz04}
  to our effective $T^2/\Z2$ orbifold model. 

  \subsection{Anomalies and the Green--Schwarz Mechanism}
    Gauge anomalies require chiral fermions\footnote{ Also (anti)self-dual tensor
     fields can contribute. Since in our model there is one tensor field of each
    type, their effects cancel.}, so they can occur in any even dimension. Gravitational 
    anomalies\footnote{Anomalies in local
     Lorentz transformations and in general coordinate transformations are equivalent in the sense
      that the anomaly can be shifted from one to the other by local
      counterterms. We will consider anomalies 
      in local Lorentz transformations and refer to those as gravitational.}, on the other hand, only
    arise in $4k+2$ dimensions (\mbox{$k=0,1,\ldots$}), hence they will appear in the 
    bulk theory, but not on the branes. 

    The anomaly $\mathcal{A}$ is defined as the (nonvanishing) gauge variation of
    the effective action, $\mathcal{A}\!\left(\Lambda\right)=\delta_\Lambda \Gamma$. 
    It can be computed from the anomaly polynomial, a (formal) closed and gauge invariant
    $\left(d+2\right)$-form $I_{d+2}$, via the Stora--Zumino descent equations
    \cite{sto83},
    \begin{align}
      \label{eq:descent}
      \mathcal{A}\!\left(\Lambda\right) \propto \int I_d^{(1)}\,, \quad
      \d I_d^{(1)}=\delta_\Lambda I_{d+1}^{(0)} \,, \quad \d I_{d+1}^{(0)}=I_{d+2}\,,
    \end{align}
    where the superscript indicates the order in the parameter
    $\Lambda$. $I_{d+2}$ is a polynomial in traces of powers of the Riemann and Yang--Mills field
    strength tensors $R$ and $F_I$, interpreted as matrix-valued two-forms $\frac{1}{2} R_{\mu\nu\,
      a}^{\phantom{\mu\nu a}b} \d x^\mu \d x^\nu$ and \mbox{$\frac{1}{2} F_{I\,
        \mu\nu\,i}^{\phantom{I\,\mu\nu\,i}j} \d x^\mu \d x^\nu$}. They are
    derived from spin and gauge 
    connection one-forms as $R=\d \Omega + \Omega^2$ and \mbox{$F_I=\d A_I +A_I^2$}, where $I$ labels the
    factors of the gauge group. Here $a,b$ are indices in the vector representation of
    $SO\left(1,d-1\right)$, $i,j$ are indices of some representation of the
    gauge group, and wedge products of forms are understood. Expressions of the
    form $\tr F_I^n$ or $\tr R^n$, the building blocks of $I_{d+2}$, are always
    closed and gauge invariant. Their coefficients in the anomaly 
    polynomial depend on the numbers, representations and charges of the fermions under the
    respective gauge groups. 
    
    For the Green--Schwarz mechanism to cancel the anomalies, we exploit the transformation
    properties of the two-form $B_2=\frac{1}{2}B_{\mu\nu}\d x^\mu \d x^\nu$. Its variation under
    gauge and Lorentz transformations with parameters $\Lambda_I$ and $\Theta$ is
    \begin{align}
      \delta B_2 &= 
      \tr \left( \Theta \d \Omega \right) 
      -\sum_I \alpha_I
      \tr\!\left(\Lambda_I \d A_I\right) 
      \,.
    \end{align}
    The coefficients $\alpha_I$ are \mbox{$\alpha_{\SU N}=2$} and $\alpha_{\SO N}=1$
    (the $U(1)$ coefficients are normalization dependent). The crucial feature of this
    transformation is that $\delta B_2$ itself is the descent of the closed and gauge invariant
    four-form
    \begin{align}
      X_4&=\tr R^2 -\sum_I \alpha_I \tr F_I^2\;,
    \end{align}
     such that the 3-form field strength $H_3 = {\rm d} B_2 - X_3^{(0)}$ associated with $B_2$ is
     invariant. By adding appropriate interaction terms of the $B$-field to the action it is now
     possible to achieve a complete cancellation of the reducible anomalies. 

     For $T^2/\Z2$ orbifolds, the total anomaly polynomial $I_8$ is of the form
     \begin{align}
       \label{eq:I8tot}
       I_8 &= \frac 12 I_8^{\text{bulk}} + \sum_f I_6^f
       \delta^2\!\left(y-y_f\right)\d y^5 \d y^6 \; , 
     \end{align}
     where $I_8^{\text{bulk}}$ is the anomaly polynomial on $\mathbbm{R}^{1,3} \times T^2$, and
     $I_6^f$ is the local anomaly polynomial at the  
     fixed point $f$. $I_6^f$ receives two kinds of contributions:
     Brane-localized fields and bulk fields surviving the orbifold projection at
     this particular fixed point. The latter, however, contribute with a factor
     of $\frac{1}{4}$ because the orbifold contains four fixed points. The 
     factor $\frac 12$ in (\ref{eq:I8tot}) enters since the fundamental
     domain of the orbifold is half the one of the torus. These anomalies can be cancelled by the
     Green--Schwarz mechanism if $I_8$ is reducible, i.e., if it factorizes into a product involving
     $X_4$. For the components this means 
     \begin{align} \label{eq:Ireducible}      
       I_8^{\text{bulk}}  &= \beta \,  X_4 Y_4  \,,&   I_6^f  &= \alpha \,  X_4^f
       Y_2^f \,. 
     \end{align}
     Here $X_4^f$ follows from $X_4$ by projection onto the local gauge group,
     and we have pulled out factors $\alpha = \frac{i}{48 (2 \pi)^3}$ and $\beta
     = \frac{-i}{16 (2 \pi)^3}$. Since $\tr R=\tr F=0$ for non-Abelian gauge
     groups, the localized two-forms $Y_2^f$ can only be linear combinations of
     $\U1$ field strengths, which can be redefined as $Y_2^f=c^f F^f=
     c^f \d A^f$. $A^f$ and the corresponding generator are
     referred to as the anomalous $\U1$ at the fixed point $f$. 
     
     If the anomaly polynomial factorizes in the required way, the total anomaly
     $\mathcal{A}=\int I_6^{(1)}$ descends from (\ref{eq:I8tot}) and is
     cancelled by variation of the Green--Schwarz action \cite{lnz04},
     \begin{align}
       \label{eq:GSterm}
       \begin{split}
         S_{\text{GS}} &=  \text{\Large $\int$} \left\{   -\left( \frac \beta 2 Y_3^{(0)} 
             + \alpha \sum_f c^f A^f  \delta^2\!\left(y-y_f\right)\d y^5 \d y^6 \right)
           \d B \right.\\ 
         & \left.\mspace{63mu} + \left(  \frac \beta 4  Y_3^{(0)} + \frac{\alpha}{3} \sum_f c^f
             A^f 
             \delta^2\!\left(y-y_f\right)\d y^5 \d y^6 \right) X_3^{(0)} \right\} \; .
       \end{split}
     \end{align}

  \subsection{Bulk Anomalies}
    We now check the cancellation of bulk anomalies in the model at hand. 
    It is convenient to split the gauge group index as $I=(A,u)$, with
    $A,B,\ldots$ running over the non-Abelian factors, i.e.\ $\SU6$, $\SU3$ and
    $\SO8$, while $u,v,\ldots=1,\ldots,5$ label the $\U1$ factors. The anomaly
    polynomial for the  six-dimensional case is given in Ref.~\cite{erl94}. Here
    we first check that the irreducible pieces cancel and then show 
    that the remaining parts factorize as in (\ref{eq:Ireducible}).
    
    There are three contributions in the anomaly polynomial which cannot be reducible:
    \begin{itemize}
      \item The most severe constraint arises from the quartic pure
        gravitational anomaly. The corresponding term in the anomaly polynomial
        is
        \begin{align}
          \left(244+ y -s\right)\tr R^4\,.
        \end{align}
        It is sensitive only to the number of gauginos $y$ and hyperinos
        $s$, which contribute with opposite signs due to their opposite chiralities,
        and the gravitino and dilatino. The necessary condition $s-y=244$ is
        fulfilled in our model, as can be seen from 
        Tables~\ref{tab:torus_untwisted} and \ref{tab:torus_twisted}. 
      \item Quartic non-Abelian anomalies receive contributions from the gaugino in the adjoint
        representation which need to be cancelled by opposite-chirality hyperinos. Denoting the
        number of hypermultiplets in representation $\boldsymbol{r}^i$ of group factor $G_A$ by
        $s^i_A$, the quartic terms are
        \begin{align}
         \Tr F_A^4 -\sum_i s^i_A \tr_{\boldsymbol{r}^i} F_A^4\,,
         \quad A=\SU6,\SU3,\SO8 \;.
        \end{align}
        Here $\Tr$ and $\tr_{\boldsymbol{r}^i}$ denote traces in the adjoint
        representation and in the representation $\boldsymbol{r}^i$, respectively.
        We can convert all traces to the fundamental representation (denoted simply by
        $\tr$), which will introduce representation indices, and possibly terms \mbox{$\sim\left(\tr
            F_A^2\right)^2$}, and finally  leads to the following constraints: 
        \begin{subequations}
          \begin{align}
              &\SU6: \qquad 12+6 s^{\mathbf{20}} - s^{\mathbf{6}} -s^{\bar{\mathbf{6}}} =0 \;,\\
              &\SO8: \qquad \frac{1}{2}s^{\mathbf{8}_s}+
              \frac{1}{2}s^{\mathbf{8}_c} -s^{\mathbf{8}} =0 \;.
          \end{align}
        \end{subequations}
        $\SU3$ does not have a fourth-order Casimir invariant and hence $\tr
        F_{\SU3}^4$ does not give a condition at this point.
      \item Finally, the $\left(\text{non-Abelian}\right)^3$-Abelian anomaly has to vanish for
        reducibility. Again we convert all traces to the fundamental representation, and have to
        consider the $\U1$ charges of the hypermultiplets. We get two nontrivial conditions for
        each $\U1$ ($\SO8$ has no third-order Casimir):
         \begin{subequations}
          \begin{align}
              &\SU6: \qquad \sum_{\mathbf{6}_i}q^{\mathbf{6}_i}_u -
              \sum_{\bar{\mathbf{6}}_i}q^{\bar{\mathbf{6}}_i}_u =0\;, \\
              &\SU3: \qquad \sum_{\mathbf{3}_i}q^{\mathbf{3}_i}_u -
              \sum_{\bar{\mathbf{3}}_i}q^{\bar{\mathbf{3}}_i}_u =0\;.
          \end{align}
        \end{subequations}
        From the $\U1$ charges in Table~\ref{tab:torus_twisted} we see that also
        these constraints are satisfied. 
    \end{itemize}

    For the remaining anomaly polynomial we normalize the $\U1$'s from Table~\ref{tab:u1defs} by
    introducing $\hat{t}_u = t_u/\sqrt{2} |t_u| $. As shown in Appendix~\ref{sec:ancoeffs},
    this leads to a factorization of the bulk anomaly polynomial which is of the form
    (\ref{eq:Ireducible}): 
    \begin{align}
      \begin{split}
        \I \left(2\pi\right)^3 I_8^{\text{bulk}} &= \frac{1}{16} \left[\tr R^2 - 2 \tr F_{SU(6)}^2
          - 2 \tr F_{SU(3)}^2 - \tr F_{SO(8)}^2 - \sum_u F_u^2\right]\\
        &\mspace{25mu} \times \left[\tr R^2 - \sum_{u,v}\beta_{uv} F_u F_v\right]\\
        &=\frac{1}{16} X_4 \, Y_4\;.
      \end{split}
    \end{align}
    The symmetric coefficient matrix $\beta_{uv}$ in the  $\hat{t}_u$ basis is
    \begin{align}
      \label{eq:beta}
      \beta_{uv}&=  \begin{pmatrix}3 & -1&0 &-1 &0\\ & 3 & 0&-1& 0\\ & &2
        &0&\sqrt{2}\\ &&&4&0\\ &&&&4 \end{pmatrix} \; .
    \end{align} 
    We conclude that all bulk anomalies of our orbifold model are cancelled by variations of the
    terms $\sim Y_3^{(0)}$ in Eq.~(\ref{eq:GSterm}).

  \subsection{Brane Anomalies}
    Since our model contains one Wilson line in the $\SO4$ plane, the spectra at the fixed points
    only depend on $n_2$ and not on $n_2'$, so that we have to evaluate two anomaly polynomials
    $I_6^{0,1}$  in the following. 

    At a fixed point, there are no gravitational anomalies, and so the only 
irreducible
    contributions are non-Abelian cubic ones. Matter now comes in chiral
    multiplets which can have both chiralities and thus contribute with 
opposite
    signs. Furthermore, the anomaly  induced by bulk fields surviving the
    projection is suppressed by a factor of $\frac{1}{4}$ with respect to the
    contributions from localized fields. Taking this into account, the cubic
    non-Abelian anomalies are of the form  
    \begin{equation}
      \frac{1}{4} \sum_{\text{bulk } \bf r} \left( s_A^{ (+)  {\bf r}} - s_A^{
          (-) {\bf r}} \right) \tr_{\bf r} F_A^3 - \sum_{\text{loc } \bf r}s_A^{
        (-) {\bf r}} \tr_{\bf r} F_A^3 \;,  
    \end{equation}
    where the sum is over representations ${\bf r}$ of the local group factor
    $A$, and the $s_A^{ (+) {\bf r}}$ and $s_A^{ (-) {\bf r}}$ denote the number
    of multiplets in that representation with 
    positive and negative chirality, respectively.  We take the localized fields
    to be left-handed. Using Tables~\ref{tab:states_T2T4_nn0} to
    \ref{tab:states_T1T3T5_nn1}, one finds that the model contains no
    irreducible local anomalies. Vector 
    multiplets do not contribute to anomalies, as they are in a real
    representation of the gauge group, and neither do the hypermultiplet
    remnants of 6D vector multiplets, since they come in left- and right-handed form.

    For the local reducible anomalies we find the following
    factorization at $n_2=0,1$ (cf.~Appendix~\ref{sec:ancoeffs}):
    \begin{align}
      \begin{split}\label{eq:I60} 
        i (2 \pi)^3 I_6^{0}=& -\frac{1}{48} \Big[ \left( \tr R^{2\vphantom{'}} \right) - 2
        \left( \tr F_{\SU5\vphantom{'}}^2 \right) - 2 \left( \tr F_{\SU3\vphantom{'}}^2 \right) \Big. \\ 
        & \Big.\mspace{100mu}-  \left( \tr F_{\SO8\vphantom{'}}^2 \right)  -  \sum_{u=1}^6
        F_u^2 \Big]  
        \times \left( \tr_0 \hat{t}_{\rm an}^0 \right) F^0 \,,
      \end{split}\\
      \begin{split}
        i (2 \pi)^3 I_6^{1}=& -\frac{1}{48} \Big[ \left( \tr R^{2\vphantom{'}} \right) - 2
        \left( \tr F_{\SU2\vphantom{'}}^2 \right) - 2 \left( \tr F_{\SU4\vphantom{'}}^2 \right) - 2
        \left( \tr F_{\SU2'}^2 \right)\Big. \\  
        & \Big. \mspace{100mu} - 2 \left( \tr
          F_{\SU4'}^2 \right)  
        -  \sum_{v=1}^8 F_v^2 \Big] \times  \left( \tr_1 \hat{t}^1 \right)
        F^1\,. 
        \label{eq:I61} 
      \end{split}
    \end{align}
    The traces of the anomalous $\U1$'s are the sums of the charges of the
    fields present at the given fixed point, and again the contributions of
    surviving bulk fields are weighted with a factor of $\frac{1}{4}$. The
    indices $u,v$ in the formulae above run over a basis spanned by the
    anomalous  $\U1$ and orthogonal generators, 
$\hat{t}_1^f \equiv \hat{t}_{\rm an}^f$,
$\hat{t}_{\rm an}^f \cdot \hat{t}_u^f = 0$, $(u>1)$. The normalization is chosen such
    that all Abelian factors have 
    level 1, namely $\hat{t}_u^f = t_u^f/\sqrt{2} |t_u^f |$. The
    factorization is of the form (\ref{eq:Ireducible}) such that we conclude that all
    anomalies of our model are cancelled by the localized part of the 
    Green--Schwarz term (\ref{eq:GSterm}). 

    Equations (\ref{eq:I60}) and (\ref{eq:I61}) reveal that due to the presence
    of one Wilson line there are two distinct anomalous $\U1$ factors $
    t_{\rm an}^0$ and $t_{\rm an}^1$ in the model, one for each
    inequivalent fixed point. For the (unnormalized) anomalous generators from
    Table~\ref{tab:u1defs} we find the following traces:   
    \begin{equation}
      \tr_0 t_{\rm an}^0 = 2 |t_{\rm an}^0|^2 = 148 \;, \quad
      \tr_1 t_{\rm an}^1 =   |t_{\rm an}^1|^2 = 80 \;. \label{eq:antr}
    \end{equation}
    The 4D anomalous $\U1$ follows from integrating the
    Green--Schwarz term over the internal dimensions. As can be seen
    from~(\ref{eq:GSterm}), this amounts to summing the normalized local
    $\U1$'s. The four-dimensional anomaly polynomial again is of the form 
    (\ref{eq:Ireducible}), so we can deduce the anomalous $\U1$ in four dimensions
    from 
    \begin{equation}
      \label{eq:tan4draw}
      \frac{\tr_\text{4d} t_{\rm an}^{(\text{4d})}}{|t_{\rm an}^{(\text{4d})}|^2}
      \ t_{\rm an}^{(\text{4d})} =  
      2 \left( \frac{\tr_0 t_{\rm an}^0}{|t_{\rm an}^0|^2} \ t_{\rm an}^0
        + \frac{\tr_1 t_{\rm an}^1}{|t_{\rm an}^1|^2} \ t_{\rm an}^1 \right)\;.
    \end{equation}
    Here $\tr_\text{4d}$ denotes the trace over the low-energy spectrum,
    i.e.~zero modes of bulk fields and localized fields, but excluding bulk
    fields which only survive at $n_2=0$ or $n_2=1$. Note that the factor of
    $\frac{1}{4}$ included in the definitions of $\tr_0$ and $\tr_1$ ensures
    that zero mode contributions are counted only once. Thus we find the
    anomalous generator $\hat{t}_{\rm an}^{(\text{4d})}$ from \cite{bhx06} with
    $\tr t_{\rm an}^{(\text{4d})} = 12 \ |t_{\rm an}^{(\text{4d})}|^2 = 88$ as 
    \begin{equation}
    t_{\rm an}^{(\text{4d})} = \frac16 \left( 2 \ t_{\rm an}^0 + t_{\rm an}^1 
      \right) \; . \label{eq:tan4d} 
    \end{equation}

    So all appearing anomalies have been cancelled, either among themselves or
    by the Green--Schwarz mechanism. We would like to emphasize that there is no
    free parameter involved: the fields and gauge groups are fixed, as well as
    the transformation property of $B_{MN}$, which is the only available
    antisymmetric tensor field which can cancel anomalies. Hence the way in
    which the different sectors combine in the correct way appears highly 
    non-trivial.  


\section{Decoupling of Exotic States}
\label{sec:decoupling}

Let us now consider the decoupling of states with exotic standard model
quantum numbers. These are the $\SU5$ $\F$-plets of bulk hypermultiplets 
which originate from the $T_2/T_4$-- and the untwisted sector, and the $\SU2$
doublets and singlets with non-zero hypercharge from the $T_1/T_5$-- and 
$T_3$--sectors at the fixed points $n_2=1$. Note that no exotic matter is 
located at the fixed points $n_2=0$. All the exotic $\F$-plets and most of 
the exotic matter at $n_2=1$ can be decoupled by VEVs of just a few standard 
model 
singlet fields. This decoupling takes place locally at one of the fixed 
points, which is a crucial difference compared to previous discussions of 
decoupling in four dimensions \cite{bhx06,lnx06}. 

The $\mathcal{N}=2$ hypermultiplets $H=(H_L,H_R)$ consist of pairs of 
$\mathcal{N}=1$ left- and right-chiral multiplets either from the $T_2$ 
and $T_4$ twisted sectors, or from the untwisted sector. The charge 
conjugate left-chiral multiplet 
$H_R^c$ has the opposite gauge quantum numbers as $H_L$. Hence the $\SU5$ 
$\F$- and $\Fb$-hypermultiplets contain the exotic $\mathcal{N}=1$ 
left-chiral multiplets $\F$ and $\Fb^c$.

The products $\F_{n_3}\F^c_{n_3}$ and $\Fb_{n_3}\Fb_{n_3}^c$, 
$n_3=0,1,2$, are total gauge singlet 
$\mathcal{N}=1$ chiral multiplets. They do carry, however, non-zero 
$R$-charges, $R=(-1,-1,0)$ (cf.~App.~\ref{ssec:Rcharges}). 
One easily verifies 
(cf.~Tables~\ref{tab:Ri}, \ref{tab:states_T2T4_nn0} and
\ref{tab:states_T1T3T5_nn0}) 
that the product $\bar{Y}_0^cS_1S_5$ of standard model singlet fields is a 
total gauge singlet with $R$-charges \mbox{$R=(0,0,-1)$}. $S_1$ and $S_5$ are oscillator
states localized at the fixed points $n_2=0$. One therefore obtains the local 
$\mathcal{N}=1$ superpotential terms
\begin{equation}\label{w1}
  W_1 = \bar{Y}_0^c S_1 S_5\left(\F_0\F_0^c + \Fb_0\Fb_0^c + \F_1\F_1^c 
  + \Fb_1\Fb_1^c + \F_2\F_2^c + \Fb_2\Fb_2^c \right)\;.
\end{equation} 
All terms are total gauge singlets with $R$-charges $R=(-1,-1,-1)$. Hence, the
H-momentum rules are satisfied, as are the space selection rules 
(cf.~\cite{bhx06}).

\begin{table}[t]
  \begin{center}
    \begin{tabular}{|c|ccc|ccccc|}\hline
      & $\F$ & $\Fb_0^c$ & $\F_1$& $\Fb$& $\F_0^c$ & $\Fb_1$& $\Fb_2$     &
      $\F_2^c$\jvs\\\hline\hline
      $\SU3\times \SU2$ & $\left(1,\mathbf{2}\right)$ & 
      $\left(\mathbf{3},1\right)$ & $\left(1,\mathbf{2}\right)$ & 
      $\left(1,\mathbf{2}\right)$ & $\left(\bar{\mathbf{3}},1\right)$ &
      $\left(1,\mathbf{2}\right)$ & $\left(\bar{\mathbf{3}},1\right)$ & 
      $\left(1,\mathbf{2}\right) $ \jvs\\\hline
      $\U1_{B-L}$ & 0 & $-\frac{2}{3}$ & 0 & 0 & $\frac{2}{3}$ &  0 & 
      $ -\frac{1}{3}$ & $-1$ \jvs\\\hline 
      MSSM  & $H_u$ & & & & & $H_d$ & $d_3$ & $l_3$ \jvs\\\hline 
    \end{tabular}
    \caption{The remaining $\F$'s and $\Fb$'s after the decoupling through 
    $W_1$. The $\SU3\times\SU2$ representations, $B-L$ charges and MSSM 
    identification refer to the zero modes. \label{tab:5decopling}}
  \end{center}
\end{table}


From Eq.~(\ref{w1}) we conclude that a large vacuum expectation value 
$\left<\bar{Y}_0^cS_1S_5\right>$ removes 6 pairs of $(\F,\Fb)$-plets\footnote{
When the distinction between $T_2$--, $T_4$-- and untwisted sector does not 
matter, we collectively denote $\F$ and $\Fb^c$ by $\F$, and $\Fb$ and
$\F^c$ by $\Fb$.}  
from the low energy spectrum. Since we have 3 positive parities for each
value of $n_3$ (cf.~Tables 3.2 and A.3), 6 $\F$- or $\Fb$-plets remain.
The mass terms are localized at the fixed points $n_2=0$. Bulk mass terms 
between hypermultiplets are forbidden by $\mathcal{N}=2$ supersymmetry. 

Inspection of Tables~\ref{tab:states_A_nn0} and  \ref{tab:states_T2T4_nn0} 
shows that from the $T_2$-, $T_4$- and untwisted sectors three $\F$'s
and five $\Fb$'s  remain: $\F$, $\Fb$, $\F^c_0$, $\Fb^c_0$, $\F_1$, $\Fb_1$, 
$\F^c_2$, $\Fb_2$. The further
decoupling is motivated by phenomenological arguments and by simplicity.
The projection condition at the fixed points $n_2=1$ leave as 4D zero
modes from each $\F$ and $\Fb$ either an $\SU3$ triplet or an $\SU2$
doublet. With respect to the $\U1_{B-L}$ generator identified in
\cite{bhx06},
\begin{equation}
 t_{B-L} = 
 \left(0,1,1,0,0,-\frac{2}{3},-\frac{2}{3},-\frac{2}{3}\right)\,
 \left(\frac{1}{2},\frac{1}{2},\frac{1}{2},-\frac{1}{2},0,0,0,0\right)\;,
\end{equation}
these massless states have the $B-L$ charges listed in 
Table~\ref{tab:5decopling}. This
suggests to decouple $\Fb_0^c$ and $\F_0^c$, which is possible with
a local coupling at the fixed point $n_2=0$,
\begin{equation}\label{w2}
W_2 = Y_0 S_1 S_5\ \F_0^c \Fb_0^c \;, 
\end{equation}
and a large VEV $\langle Y_0 S_1 S_5 \rangle$.

From the remaining $\F$--plets,
either $\F$ or $\F_1$ can be chosen as Higgs field $H_u$. A large top-quark 
coupling is obtained for $\F \supset H_u$. $\F_1$ can be easily decoupled 
using the 6D gauge coupling with the chiral multiplet $\Fb$ of the 
$\SU6$ ${\bf 35}$-plet,
\begin{equation}
W_H \supset \sqrt{2}g \left(X_0 \F\F_0^c + \bar{X}_0 \Fb\Fb_0^c + 
X_1^c \F_1\Fb +  \bar{X}_1^c \Fb_1\F + X_2 \F\F_2^c 
+ \bar{X}_2^c \Fb_2\F  \right)\;, 
\end{equation} 
with a large VEV $\langle\bar{X}_1^c\rangle$. The remaining $\Fb$--plets
$\F_2^c$ and $\Fb_1$ then correspond to a lepton doublet and the Higgs
field $H_d$, respectively. The chosen vacuum is similar to the 
$B-L$ conserving vacuum discussed in \cite{bhx06}. It corresponds to partial 
gauge-Higgs unification for $H_u$. If one chooses to decouple $\F$ instead 
of $\F_1$, one has no gauge-Higgs unification. Alternatively, one can also 
keep $\F$ and $\Fb$ massless, corresponding to full gauge-Higgs unification.

All other exotic states are localized at $n_2=1$. The $\SU2$ doublets $M_i$
and some of the $\SU2$ singlets $S^{\pm}_i$ can already be decoupled by
cubic terms,
\begin{align}
  W_3 &= \bar{Z}_1^c M_1 M_4 + Z_0^c M_2 M_3\;, \label{w3} \\
  \begin{split}
    W_4 &= \bar{Y}_2^c \left(S_2^+ S_1^- + S_3^+ S_4^- \right) +  
    Z_2 \left(S_4^+ S_5^- + S_4^+ S_5^{'-} \right)\\
    & \hspace{0.3cm} +  \bar{Z}_2 \left(S_3^- S_6^+ + S_3^- S_6^{'+} \right)
    + U_1^c \left( S_6^+ S_5^- + S_6^{'+} S_5^{'-} \right) \; ,\label{w4}
  \end{split}
\end{align}
with large VEVs $\langle\bar{Z}_1^c\rangle$, $\langle Z_0^c\rangle$,
$\langle\bar{Y}_2^c\rangle$, $\langle Z_2\rangle$, $\langle U_1^c\rangle$.
The decoupling of the remaining exotic singlets with hypercharge, $S_1^+$,
$S_2^-$, $S_5^+$, $S_6^-$, $S_7^-$, $S_7^+$ requires higher dimensional
operators (cf.~\cite{bhx06,lnx06}), which we will not discuss further in 
this paper.

After the decoupling of altogether 8 pairs of $(\F,\Fb)$--plets we are left
with two localized families,
\begin{equation}
(n_2,n_2') = (0,0):\ \Fb_{(1)},\; \T_{(1)};\quad   
(n_2,n_2') = (0,1):\ \Fb_{(2)},\; \T_{(2)}\;,
\end{equation}
together with two further families and a pair of Higgs doublets in the bulk:
\begin{equation}
\Fb_{(3)} \equiv \F^c_2,\;  \T_{(3)} \equiv \T; \quad 
\Fb_{(4)} \equiv \Fb_2,\;  \T_{(4)} \equiv \Tb^c; 
\quad H_u \equiv \F,\; H_d \equiv \Fb_1\;.
\end{equation}
At the fixed points $n_2 = 0$ these chiral $\mathcal{N} = 1$ multiplets
form a local $\SU5$ GUT theory. The corresponding Yukawa couplings will be
discussed in the following section. From the two bulk families, half of the states are projected out
by the projection conditions at $n_2=1$, and together they give rise to one family of zero modes
(cf.~Eq.~(\ref{eq:qe10}) and  Tab.~\ref{tab:5decopling}). 

Note that the decoupling terms (\ref{w1}), (\ref{w2}), (\ref{w3}) and 
(\ref{w4}) require VEVs of both bulk and localized fields. The
localized singlets $S_1$ and $S_5$ correspond to oscillator modes.
As we will see in Section~\ref{sec:vacuum}, bulk and brane field backgrounds
are typically induced by local Fayet--Iliopoulos (FI) terms. The non-vanishing 
VEVs of localized fields are often related to a resolution of the orbifold 
singularities \cite{fix88,gnx07}. However, a study of the blow-up of
the considered orbifold to a smooth manifold and the geometrical 
interpretation of the localized VEVs is beyond the scope of this work.


\section{Yukawa Couplings}
In the previous section we have obtained four quark-lepton
families transforming as \mbox{($\Fb_{(i)}+\T_{(i)}$)} under $\SU5$, where $i$ is a 
generation index. Two families are localized at the branes ($i=1,2$) and
two are bulk fields. The corresponding superpotential reads
\begin{align}\label{yuk5}
  W_\mathrm{Yuk} =  
  C_{ij}^{(u)}\T_{(i)}\T_{(j)} H_u + C_{ij}^{(d)}\Fb_{(i)}\T_{(j)}H_d\;,
\end{align}
where the couplings $C_{ij}^{(u)}$ and $C_{ij}^{(u)}$ are composed of singlet fields such that the
superpotential obeys the string selection rules (cf.~\cite{bhx06}).

As an example, we consider a vacuum where in addition to the fields
\begin{equation}
\bar{Y}^c_0, S_1, S_5, Y_0, X_1^c, \bar{Z}^c_1, Z^c_0, \bar{Y}^c_2,
Z_2, \bar{Z}_2, U^c_1\;,
\end{equation}
used in Section~5 for decoupling, only the singlets 
\begin{equation}
  Y_0^c, Y_1, \bar{Y}_1, S_3, S_4, S_7
\end{equation}
acquire non-zero VEVs. After a straightforward calculation, we find that
up to $\mathcal{O}(8)$ in the fields, this vacuum leads to couplings
\begin{align}
C_{ij}^{(u)} &= \left( \begin{array}{cccc}
a_1 & 0 & a_2 & a_3 \\
0 & a_1 & a_2 & a_3 \\
a_2 & a_2 & 0 & g \\
a_3 & a_3 & g & a_4
\end{array} \right) , &
C_{ij}^{(d)} &= \left( \begin{array}{cccc}
0 & 0 & b_1 & b_2 \\
0 & 0 & b_1 & b_2 \\
b_3 & b_3 & b_4 & 0 \\
b_5 & b_5 & b_6 & b_5^2
\end{array} \right), 
\end{align}
with 
\begin{align}
  a_1 &= \langle Y_0^c \bar{Y}_0^c S_1 S_3 \rangle ,&
  a_2 &= \langle \left( \bar{Y}_0^c S_1 \right)^2 S_5 \rangle, &
  a_3 &= \langle Y_0^c \bar{Y}_0^c S_1 S_3 S_5 \rangle, \\
  a_4 &= \langle Y_0^c \bar{Y}_0^c S_1 S_3 \left( S_5 \right)^2 \rangle, \\
  b_1 &= \langle Y_0 \bar{Y}_1 \left( S_5 \right)^3 \left( S_7 \right)^2 \rangle, &
  b_2 &= \langle X_1^c \bar{Y}_2^c U_1^c S_7 \rangle, &
  b_3 &= \langle X_1^c \bar{Y}_1 S_3 \left( S_5 S_7 \right)^2 \rangle, \\
  b_4 &= \langle \left( X_1^c \right)^2 \bar{Y}_1 U_1^c S_4 S_7 \rangle, &
  b_5 &= \langle S_5 \rangle, &
  b_6 &= \langle \left( X_1^c \right)^2 Y_1 S_1 S_7 \rangle \;.
\end{align}
Note that the chosen vacuum yields non-vanishing Yukawa couplings while the 
$\mu$-term is only generated at higher order.

The Yukawa couplings (\ref{yuk5}) are $\SU5$ invariant, hence we have obtained
an $\SU5$ GUT model. Note that the SU(5) Yukawa interactions are local since 
the fields $S_i$ are localized at the fixed points $n_2 = 0$, i.e., we have a
{\it local} $\SU5$ GUT model. The only exception is 
$C_{34}^{(u)}=C_{43}^{(u)}=g$, which is a remnant of the SU(6) bulk gauge 
interaction. It is a consequence of the partial gauge-Higgs unification
of the present model, which implies a phenomenologically attractive 
large top Yukawa coupling.

\begin{figure}[t]
  \begin{center}
    \begin{picture}(250,200)\small
      \SetOffset(25,25)
      \CArc(-172.5,75)(205,-25,25)
      \CArc(372.5,75)(205,155,205)
      \CArc(100,-197.5)(205,65,115)
      \CArc(100,347.5)(205,245,295)
      {\SetColor{Blue} \Vertex(14,161){3} \Vertex(14,-11){3}}
      {\SetColor{Red} \Vertex(186,161){3} \Vertex(186,-11){3}} 
      \Text(5,161)[r]{$\Fb\oplus\T$} \Text(5,-11)[r]{$\Fb\oplus\T$} 
      \Text(100,82)[]{$2\times \Fb$, $2\times \T$}
      \Text(100,68)[]{$\F$, $\Fb$}
      \LongArrow(80,90)(24,151)
      \LongArrow(80,60)(24,-1)
      \Text(195,161)[l]{exotics}  \Text(195,-11)[l]{exotics}     
    \end{picture}
  \end{center}
  \caption{The orbifold $T^2/\Z2$. The blue dots (on the left) label the fixed points with $n_2=0$,
    the red ones (right) have $n_2=1$. Two quark-lepton generations live at the $n_2=0$ fixed points, the
    third one originates from two $\SU5$ $\Fb$ and $\T$ multiplets in the bulk, half of which is projected out
    due to the boundary conditions at $n_2=1$. \label{fig:t2z2orbi}}
\end{figure}
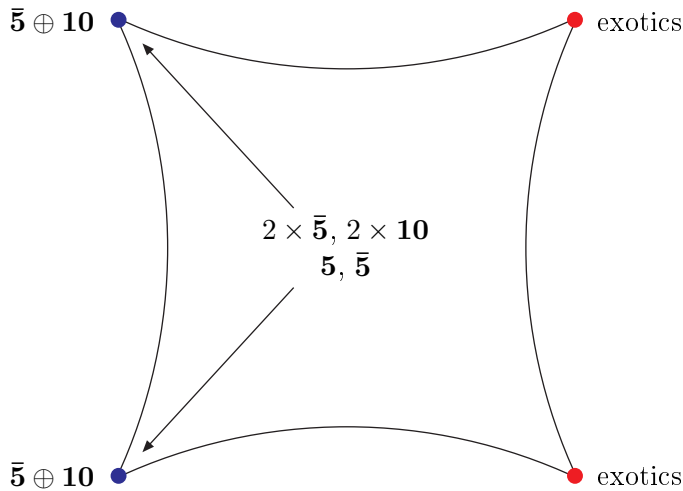

We can now proceed and deduce the corresponding Yukawa couplings in four 
dimensions. As described in Section~\ref{sec:decoupling}, half of each of
 the two bulk families is projected out by the additional $\Z2$ orbifold 
condition at the second pair of fixed points ($n_2=1$).
 The remaining fields from the split bulk matter multiplets then form the content of the third
 standard model family. The 4D Yukawa terms are 
 \begin{align}
  W_\mathrm{Yuk} &= Y_{ij}^{(u)} u_i^c q_j H_u + Y_{ij}^{(d)} d_i^c q_j H_d +  
  Y_{ij}^{(l)} l_i e_j^c H_d\;,
 \end{align}
 where $i,j=1,2,3$ is a family index, and
 \begin{align}
   Y_{ij}^{(u)} &= \left( \begin{array}{ccc}
    a_1 & 0 & a_3 \\
    0 & a_1 & a_3 \\
    a_2 & a_2 & g
   \end{array} \right), &
   Y_{ij}^{(d)} &= \left( \begin{array}{ccc}
    0 & 0 & b_2 \\
    0 & 0 & b_2 \\
    b_5 & b_5 & b_7
   \end{array} \right), &
   Y_{ij}^{(l)} &= \left( \begin{array}{ccc}
    0 & 0 & b_1 \\
    0 & 0 & b_1 \\
    b_3 & b_3 & b_4
   \end{array} \right).
 \end{align}
The Yukawa matrices for down quarks and leptons are different, although
they originate from $\SU5$ invariant couplings of the 6D theory. This is 
due to the split multiplets which form the third quark-lepton family.
In this way the mostly unsuccessful $\SU5$ predictions for fermion masses
are avoided. However, one also loses the successful prediction 
$m_b(M_\mathrm{GUT}) \simeq m_{\tau}(M_\mathrm{GUT})$.

The obtained local $\SU5$ GUT model is phenomenologically not viable. Not
only are electron and down-quark massless, which may be corrected by higher
powers of singlet VEVs, but the main problem are $R$-parity violating Yukawa 
couplings leading to rapid proton decay, which we have not listed.
However, the present model is just an example of a large class of models
\cite{lnx06}, and it is likely that the phenomenology can be improved.
In the above discussion we have also ignored neutrino masses which can
be generated by a seesaw mechanism typically involving many singlet fields
\cite{bhx07}.

 \section{Supersymmetric Vacua}
\label{sec:vacuum}

In the previous sections we have discussed phenomenologically wanted vacuum
configurations, i.e. expectations values of singlet fields, which decouple
states with exotic quantum numbers and generate Yukawa couplings for quarks
and leptons.
The analysis and classification of these vacua is a difficult problem. In
particular, one has to show that $\mathcal{N}=1$ supersymmetry remains unbroken
in four dimensions. For the present model the conditions for
vanishing $F$- and $D$-terms have been discussed in \cite{bhx06}. A crucial
role is played by the Fayet--Iliopoulos $D$-term of the anomalous $\U1$, which 
drives fields away from zero (cf.~\cite{ccx99}). 

In this paper we are studying the case where two of the compact dimensions are
larger than the other four. Such an ansatz assumes that the size of the large 
dimensions can be stabilized at a scale 
$1/M_\mathrm{GUT} \gg 1/M_\mathrm{string}$. To prove this one has to find
supersymmetric vacua of the effective 6D field theory which incorporates  
Kaluza--Klein states with masses between $M_\mathrm{GUT}$ and 
$M_\mathrm{string}$.

As we saw in Section~4, the 6D theory has different Fayet-Iliopoulos terms at
the inequivalent fixed points in the $\SO4$-plane (cf.~(\ref{eq:tan4draw})),
\begin{equation}
\mathcal{L}_\mathrm{FI} = \sum_f\ \xi_f\ \delta^2(y-y_f) 
\left(-D_3^f + F^f_{56}\right)\;,
\end{equation}
where at $f=(n_2,n_2')$,
\begin{equation}
\xi_{(0,0)} = \xi_{(0,1)} = 
\frac{g M_\mathrm{P}^2}{384\pi^2} \frac{ \mathrm{tr}_0 \ t^0_\mathrm{an}}{ \left| t_{\rm an}^0 \right|^2 }
\;, \quad
\xi_{(1,0)} = \xi_{(1,1)} = 
\frac{g M_\mathrm{P}^2}{384\pi^2} \frac{ \mathrm{tr}_1 \ t^1_\mathrm{an}}{ \left| t_{\rm an}^1 \right|^2 }
\;.
\end{equation}
Integrating over the two compact dimensions reproduces the 4D Fayet-Iliopoulos
term of \cite{bhx06}. 

In the case of flat space, localized FI terms have been studied in 
\cite{lnz04}, and it has been shown that they lead to an instability of the
bulk fields and to spontaneous localization towards the fixed points. 
For our 6D supergravity theory this analysis has to be extended to include
the gravitational, antisymmetric tensor and dilaton fields. In general,
one expects warped solutions, and it is not clear whether $\mathcal{N}=1$
supersymmetry remains unbroken in four dimensions. These questions are
beyond the scope of the present paper and will be studied elsewhere.

In the following we will only check whether the VEVs selected in Sections~5 
and 6 correspond to a supersymmetric vacuum for an isotropic orbifold, where
the $\SO4$-, $\SU3$- and $G_2$-planes all have string size, and the different
FI terms are approximated by a single FI term in four dimensions. As
discussed in \cite{bhx06}, vanishing $D$-terms are guaranteed if all fields
are part of gauge invariant monomials except one which carries negative
net anomalous charge. These conditions are indeed satisfied for the vacuum
chosen in Sections~5 and 6. Explicit examples of gauge invariant monomials are
\begin{equation}
X_3 X_3^c, \; X_4^c S_1 S_5, \; X_5^c X_{12}^c Y_1^c Y_4^c S_7^2 \;, 
 X_5^c X_8 Y_5 Y_6 S_4 S_7, \; X_5^c X_8 X_{12}^c Z_1^c S_3 S_7 \;,
\end{equation}
supplemented by
\begin{equation}
X_3^c (X_5^c X_7)^2 Y_8
\end{equation}
which has anomalous charge $-22/3$.  

Since the superpotential of the standard model singlet fields is unknown,
we cannot prove that the $F$-terms vanish for the chosen vacuum.We expect,
however, a simplification in the analysis of the superpotential in 6D as
compared to 4D, since the superpotential is generated locally at the fixed
points where one has larger unbroken symmetries than in the 4D effective
theory.

It will be very interesting to see whether a supersymmetric vacuum of
an isotropic orbifold can be obtained as limiting case from an anisotropic
orbifold. The different FI terms at the orbifold fixed points may play
a crucial role in generating the anisotropy, and it is intriguing that
the mass scale of the FI terms is of the order of the grand unification scale,
$M_\mathrm{P}/\sqrt{384 \pi^2} \sim M_\mathrm{GUT}$.

\section{Outlook}\label{sec:Outlook}

We have constructed a 6D supergravity theory as intermediate step in the
compactification of the heterotic string to the supersymmetric standard
model in four dimensions. The theory has $\mathcal{N}=2$ supersymmetry
and one tensor multiplet, and it has a large number of gravitational, gauge and
mixed anomalies, all of which are cancelled by the Green--Schwarz mechanism.
The theory is compactified from six to four dimensions on a $\Z2$ orbifold
with two inequivalent pairs of fixed points with unbroken $\SU5$ and
$\SU2\times\SU4$ symmetry, respectively.

In addition to the cancellation of anomalies, we have been particularly
interested in the decoupling of exotic states and the emergence of an 
intermediate $\SU5$ GUT. Compared to the 4D theory the decoupling is more
transparent due to the larger symmetries, $\mathcal{N}=2$ supersymmetry in
the bulk and larger gauge symmetries at the orbifold fixed points. It is 
remarkable that most exotic states can be decoupled with VEVs of a few
standard model singlet fields at the orbifold fixed points.

A very interesting feature of the theory is the emergence of an intermediate
$\SU5$ GUT model. Two quark-lepton families are localized at the $\SU5$ 
branes and two further families, together with a pair of $\F \oplus\Fb$ plets 
are bulk fields. $\SU5$ is broken by the presence of the $\SU2\times\SU4$
branes. This generates a pair of Higgs doublets as split multiplets. Split
multiplets of the two bulk quark-lepton families also form the third 
quark-lepton family,
with the standard model quantum numbers of one $\Fb$-plet and one $\T$-plet.
Due to the presence of the split multiplets, the Yukawa couplings of the 4D
theory break $\SU5$ explicitly, thus avoiding unsuccessful $\SU5$ predictions
of ordinary 4D GUTs.

The 6D theory originally has a large number of $\F \oplus \Fb$ pairs, most of
which are decoupled. As discussed in Section~5, the identification of the
Higgs fields depends on the choice of the vacuum configuration, and one
can have no, partial or full gauge-Higgs unification. Since there is no
clear distinction between matter and Higgs fields, one generically
expects large $R$-parity breaking Yukawa couplings leading to fast proton decay,
as it is indeed the case for the vacuum chosen in Sections~5 and 6.
However, since the considered model is just one example of a large class
of similar models \cite{lnx06}, it is likely that the phenomenology can
be improved. 

On the theoretical side, the main open problems concerns the stabilization
of extra dimensions at a scale $1/M_\mathrm{GUT} \gg 1/M_\mathrm{string}$
and the existence of corresponding vacua with unbroken $\mathcal{N}=1$
supersymmetry. We hope to address these questions elsewhere.

\vspace{1.5cm}

\noindent
\begin{minipage}{\textwidth}
{\bf\large Acknowlegments}\\

We would like to thank S. Groot Nibbelink, A. Hebecker, J. Louis and 
M. Trapletti for valuable discussions.
\end{minipage}

\clearpage
\appendix

\section{States}\label{app:states}

\subsection[$R$-Charges]{\boldmath$R$-Charges}
\label{ssec:Rcharges}
The $R$-charges of a chiral multiplet are defined as
$R^i = q_{\rm sh}^i - (\widetilde{N}-\widetilde{N}^*)^i$, where $q_{\rm sh}^i$ is the
shifted H-momentum of the scalar and the vectors $\widetilde{N}$ and $\widetilde{N}^*$ denote 
oscillator numbers of left-moving fields in $z^i$ and $\bar{z}^i$ directions, respectively.
\begin{table}[hb]
	\centering
	\begin{tabular}{|c|c|c|c c c|}
		\hhline{------}
		\jvs Sector & State & Excitation & $R^1$ & $R^2$ & $R^3$ \\
		\hhline{======}
		\jvs $U$ & $U_1^c$ & &0 & $-1$ & 0 \\
		\jvs  $U$ & $U_2, U_3, U_4$ & & $-1$ & 0 & 0 \\
		\hhline{------}
		\jvs  $T_1$ & All & & $-\frac16$ & $-\frac13$ & $-\frac12$ \\
		\hhline{------}
		\jvs $T_1^*$ & $S_1, S_2, S_7$ &  $\widetilde{N}^* = (1,0,0)$& $\frac56$ & $-\frac13$ & $-\frac12$ \\
		\jvs $T_1^*$ & $S_4, S_6$ &  $\widetilde{N}^* = (2,0,0)$& $\frac{11}{6}$ & $-\frac13$ & $-\frac12$ \\
		\jvs $T_1^*$ & $S_3, S_5$ &  $\widetilde{N}^* = (0,1,0)$& $-\frac16$ & $\frac23$ & $-\frac12$ \\
		\hhline{------}
		\jvs  $T_2$ & $H_L$ && $-\frac13$ & $-\frac23$ & 0 \\
		\hhline{------}
		\jvs $T_2^*$ & $Y_{n_3}^*$ & $\widetilde{N}=(0,1,0)$ & $-\frac13$ & $-\frac53$ & 0 \\
		\jvs $T_2^*$ & $Y_{n_3}^{'*}$ & $\widetilde{N}^*=(1,0,0)$ & $\frac23$ & $-\frac23$ & 0 \\
		\hhline{------}
		\jvs $T_3$ & All & & $-\frac12$ & 0 & $-\frac12$  \\
		\hhline{------}
		\jvs $T_4$ & $H_R^c$ & & $-\frac23$ & $-\frac13$ & 0  \\
		\hhline{------}
		\jvs  $T_4^*$ & $Y_{n_3}^{* c}$ & $\widetilde{N}^*=(0,1,0)$& $-\frac23$ & $\frac23$ & 0 \\
		\jvs  $T_4^*$ & $Y_{n_3}^{'* c}$ & $\widetilde{N}=(1,0,0)$& $-\frac53$ & $-\frac13$ & 0 \\
		\hhline{------}
	\end{tabular}
	\caption{
	$R$-charges and oscillator excitations of left-handed states. $U$ denotes the
	untwisted sector and a star represents non-vanishing oscillator numbers.
	\label{tab:Ri}
	}
\end{table}

\clearpage
\subsection{Bulk States}\label{app:bulkstates}
  Here we list the states of the effective 6D bulk theory. They are obtained 
  from the heterotic
  string by an $\mathbbm{Z}_3$ orbifold projection with one Wilson line, as described in Section~2.

  \begin{table}[h!]
    \centering
    \begin{tabular}{|c|c|c|c|c|c|c|c|}\hhline{--------}
      Multiplet & Representation & $t_1$ & $t_2$ & $t_3$ & $t_4$ & $t_5$ &
      \# \jvs\\\hhline{========}
       Graviton  &                       &       &       &       &       &       & 1\\
                 Tensor    &                       &       &       &       &       &       & 1\\
                 Hyper     &                       &       &       &       &       &       & 2\\
                \hhline{--------}
       Vector    & $\left(\boldsymbol{35};1,1\right)$ & &&&&& 35 \\
                           & $\left(1;\boldsymbol{8},1\right)$ & &&&&& 8 \\
                           & $\left(1;1,\boldsymbol{28}\right)$ & &&&&& 28 \\
                           & $5\times \left(1;1,1\right)$ & &&&&& 5 \\\hhline{--------}
      Hyper     & $\left(\boldsymbol{20};1,1\right)$ &$-\frac{1}{2}$
      &$\frac{1}{2}$&$0$&$0$&$0$& 20 \\
                           & $\left(1;1,\boldsymbol{8}\right)$ & $0$ & $0$ & $0$ & $-1$ & $0$ & 8 \\
                           & $\left(1;1,\boldsymbol{8}_s\right)$ & $0$ & $0$ & $0$ & $\frac{1}{2}$
                & $\frac{3}{2}$ & 8 \\
                           & $\left(1;1,\boldsymbol{8}_c\right)$ & $0$ & $0$ & $0$ & $\frac{1}{2}$
                & $-\frac{3}{2}$  & 8 \\
                           & $\left(1;1,1\right)$ %
                                 & $\frac{1}{2}$ & $\frac{1}{2}$ & $-3$ & $0$ & $0$&1 \\
                           & $\left(1;1,1\right)$ %
                                 & $-1$ & $-1$ & $0$ & $0$ & $0$&1 \\
                           & $\left(1;1,1\right)$ %
                                 & $1$ & $-1$ & $0$ & $0$ & $0$&1 \\
                           & $\left(1;1,1\right)$ %
                                 & $\frac{1}{2}$ & $\frac{1}{2}$ & $3$ & $0$ & $0$&1
                                 \\\hhline{--------}
    \end{tabular}
    \caption{The massless spectrum of the 6D theory arising from the untwisted
sector. There are 76 vector multiplets and 50 hypermultiplets.
 The second column refers
      to the representations with respect to $\SU6\times \SU3\times \SO8$,
      $t_1$--$t_5$ are the charges with respect to the $\U1$ factors of the 
bulk gauge group.
The first three multiplets arise from the 10D gravitational
      sector and are complete gauge singlets.} 
    \label{tab:torus_untwisted}
  \end{table}

  \begin{table}[h!]
    \centering
    \begin{tabular}{|c|c|c||c|c|c|c|c||c|}\hhline{---------}
      Sector    &  Representation & $n_3$& $t_1$ & $t_2$ & $t_3$ & $t_4$ & $t_5$ &\# \jvs\\\hhline{=========}

      $T_2$/$T_4$   &  $3\times\left(\boldsymbol{6};1,1\right) $ %
                              & 0 & $0$ & $-\frac{1}{3}$ & $1$ & $\frac{2}{3}$ & $0$& 18\jvs\\
                &            $3\times\left(\boldsymbol{\bar{6}};1,1\right)$%
                              & 0 & $0$ & $-\frac{1}{3}$ & $-1$ & $\frac{2}{3}$ & $0$ & 18 \jvs\\
                &            $3\times\left(1;1,1\right) $ %
                              & 0 & $-1$ & $-\frac{1}{3}$ & $0$ & $\frac{2}{3}$ & $0$ & 3 \jvs\\
                &            $3\times\left(1;1,1\right) $ %
                              & 0 & $1$ & $-\frac{1}{3}$ & $0$ & $\frac{2}{3}$ & $0$ & 3 \jvs\\
\hline
      $T_2$/$T_4$          &            $3\times\left(1;\boldsymbol{3},1\right) $ %
                              & 0 & $0$ & $\frac{2}{3}$ & $0$ & $-\frac{1}{3}$ & $1$& 9 \jvs\\
                &            $3\times\left(1;\boldsymbol{\bar{3}},1\right) $ %
                              & 0 & $0$ & $\frac{2}{3}$ & $0$ & $-\frac{1}{3}$ & $-1$ & 9 \jvs\\
                &            $3\times\left(1;1,\boldsymbol{8}\right) $ %
                              & 0 & $0$ & $\frac{2}{3}$ & $0$ & $-\frac{1}{3}$ & $0$ & 24 \jvs\\
\hline
     $T_2$/$T_4$$^*$  &            $6\times\left(1;1,1\right) $ %
                              & 0 & $0$ & $\frac{2}{3}$ & $0$ & $\frac{2}{3}$ & $0$ & 6 \jvs\\
\hline
\hline
       $T_2$/$T_4$         &            $3\times\left(\boldsymbol{6};1,1\right) $ %
                              & 1 & $0$ & $-\frac{1}{3}$ & $-1$ & $-\frac{1}{3}$ & $-1$& 18\jvs\\
                &            $3\times\left(\boldsymbol{\bar{6}};1,1\right)$%
                              & 1 & $\frac{1}{2}$ & $\frac{1}{6}$ & $0$ & $-\frac{1}{3}$ & $-1$ & 18 \jvs\\
                &            $3\times\left(1;1,1\right) $ %
                              & 1 & $0$ & $\frac{2}{3}$ & $-2$ & $-\frac{1}{3}$ & $-1$ & 3 \jvs\\
                &            $3\times\left(1;1,1\right) $ %
                              & 1 & $\frac{1}{2}$ & $-\frac{5}{6}$ & $1$ & $-1\frac{1}{3}$ & $-1$ & 3 \jvs\\
\hline
       $T_2$/$T_4$         &            $3\times\left(1;\boldsymbol{3},1\right) $ %
                              & 1 & $-\frac{1}{2}$ & $\frac{1}{6}$ & $1$ & $\frac{2}{3}$ & $0$& 9 \jvs\\
                &            $3\times\left(1;\boldsymbol{\bar{3}},1\right) $ %
                              & 1 & $-\frac{1}{2}$ & $\frac{1}{6}$ & $1$ & $-\frac{1}{3}$ & $1$ & 9 \jvs\\
                &            $3\times\left(1;1,\boldsymbol{8}_s\right) $ %
                              & 1 & $-\frac{1}{2}$ & $\frac{1}{6}$ & $1$ & $\frac{1}{6}$ &
                                 $\frac{1}{2}$ & 24 \jvs\\
\hline
      $T_2$/$T_4$$^*$          &            $6\times\left(1;1,1\right) $ %
                              & 1 & $-\frac{1}{2}$ & $\frac{1}{6}$ & $1$ & $-\frac{1}{3}$ & $-1$ & 6 \jvs\\
\hline
\hline
    $T_2$/$T_4$            &            $3\times\left(\boldsymbol{6};1,1\right) $ %
                              & 2 & $\frac{1}{2}$ & $\frac{1}{6}$ & $0$ & $-\frac{1}{3}$ & $1$& 18\jvs\\
                &            $3\times\left(\boldsymbol{\bar{6}};1,1\right)$%
                              & 2 & $0$ & $-\frac{1}{3}$ & $1$ & $-\frac{1}{3}$ & $1$ & 18 \jvs\\
                &            $3\times\left(1;1,1\right) $ %
                              & 2 & $\frac{1}{2}$ & $-\frac{5}{6}$ & $-1$ & $-\frac{1}{3}$ & $1$ & 3 \jvs\\
                &            $3\times\left(1;1,1\right) $ %
                              & 2 & $0$ & $\frac{2}{3}$ & $2$ & $-\frac{1}{3}$ & $1$ & 3 \jvs\\
\hline
     $T_2$/$T_4$           &            $3\times\left(1;\boldsymbol{3},1\right) $ %
                              & 2 & $-\frac{1}{2}$ & $\frac{1}{6}$ & $-1$ & $-\frac{1}{3}$ & $-1$& 9 \jvs\\
                &            $3\times\left(1;\boldsymbol{\bar{3}},1\right) $ %
                              & 2 & $-\frac{1}{2}$ & $\frac{1}{6}$ & $-1$ & $\frac{2}{3}$ & $0$ & 9 \jvs\\
                &            $3\times\left(1;1,\boldsymbol{8}_c\right) $ %
                              & 2 & $-\frac{1}{2}$ & $\frac{1}{6}$ & $-1$ & $\frac{1}{6}$ &
                                 $-\frac{1}{2}$ & 24 \jvs\\
\hline
     $T_2$/$T_4$$^*$           &            $6\times\left(1;1,1\right) $ %
                              & 2 & $-\frac{1}{2}$ & $\frac{1}{6}$ & $-1$ & $-\frac{1}{3}$ & $1$ & 6 \jvs\\
                              \hhline{---------}
    \end{tabular}
    \caption{The massless spectrum of the 6D theory arising from the $T_2$ and $T_4$ sectors. There are 270 hypermultiplets.
 The states are localised in the
      $\G2$ and $\SU3$ planes, which contain three fixed points each. 
The equivalent $\G2$ fixed points yield the multiplicity factor three, 
localization in the $\SU3$ plane is given by $n_3$. 
       $T_2$/$T_4$* states have non-vanishing
      oscillator numbers. \label{tab:torus_twisted}}
  \end{table}

  \clearpage

\subsection{States at the Fixed Points}\label{ssec:locstates}
  Here we list the states at the fixed points $n_2=0,1$.
 These involve bulk states from the $T_2/T_4$ and the untwisted sector (see
 Tables~\ref{tab:states_A_nn0} -- \ref{tab:states_BC_nn1}) and localized states from the sectors
 $T_1/T_5$ and $T_3$. $X_{i}$, $\bar{X}_{i}$, $Y_{i}$, $\bar{Y}_{i}$, $Z_{i}$, $\bar{Z}_{i}$ 
 and $U_i$ are bulk fields; $S_1$~--~$S_8$ are localized fields.

  \begin{center}
\small
\begin{longtable}{|c||c|c||c|c||c|c|c|c|c|c||c|}
\hline
\multicolumn{1}{|c||}{ Bulk } &
\multicolumn{1}{c|}{$n_2=0$} &
\multicolumn{1}{c||}{ $n_3$} &
\multicolumn{1}{c|}{$H_L$} &
\multicolumn{1}{c||}{$H_R$} &
\multicolumn{1}{c|}{$t_6^0$} &
\multicolumn{1}{c|}{$t_1$} &
\multicolumn{1}{c|}{$t_2$} &
\multicolumn{1}{c|}{$t_3$} &
\multicolumn{1}{c|}{$t_4$} &
\multicolumn{1}{c||}{$t_5$\jvs}&
\multicolumn{1}{c|}{}
 \\ \hline \hline
\endfirsthead

\multicolumn{12}{c}%
{{ \tablename\ \thetable{} -- continued from previous page}} \\
\hline
\multicolumn{1}{|c||}{ Bulk } &
\multicolumn{1}{c|}{$n_2=0$} &
\multicolumn{1}{c||}{ $n_3$} &
\multicolumn{1}{c|}{$H_L$} &
\multicolumn{1}{c||}{$H_R$} &
\multicolumn{1}{c|}{$t_6^0$} &
\multicolumn{1}{c|}{$t_1$} &
\multicolumn{1}{c|}{$t_2$} &
\multicolumn{1}{c|}{$t_3$} &
\multicolumn{1}{c|}{$t_4$} &
\multicolumn{1}{c||}{$t_5$\jvs}&
\multicolumn{1}{c|}{}
\\ \hline \hline
\endhead

\endlastfoot

\multicolumn{12}{c}%
{{ \tablename\ \thetable{} -- continued on next page}}  \\
\endfoot

\jvs  $ ({\bf 6};1,1)$ & $ ({\bf 5};1,1)$ & $0$  & $-,+,-$ & $+,-,+$ & $-1$ & 0 & $-\frac13$ & $1$ & $\frac23$ & 0 & \\
\jvs  	     & $ (1;1,1)$       & $0$  & $+,-,+$ & $-,+,-$ & $5$ & 0 & $-\frac13$ & $1$ & $\frac23$ & 0 & $X_0$\\
\hline
\jvs $ ({\bf \bar{6}};1,1)$ & $ ({\bf \bar{5}};1,1)$ & $0$  & $-,+,-$ & $+,-,+$ & 1 & 0 & $-\frac13$ & $-1$ & $\frac23$ &0 & \\
\jvs     & $ (1;1,1)$    & $0$  & $+,-,+$ & $-,+,-$ & $-5$ & 0 & $-\frac13$ & $-1$ & $\frac23$ & 0 &  $\Xb_0$\\
\hline
\jvs  $ (1;1,1)$ & $ (1;1,1)$ & $0$  & $+,-,+$ & $-,+,-$ & 0 & 1 & $-\frac13$ & 0 & $\frac23$  & 0 &  $Y_0$\\
\hline
\jvs  $ (1;1,1)$ & $ (1;1,1)$ & $0$  & $+,-,+$ & $-,+,-$ & 0 & $-1$ & $-\frac13$ & 0 & $\frac23$ & 0 &  $\Yb_0$\\
\hline
\hline
\jvs  $ (1;{\bf 3},1)$ & $ (1;{\bf 3},1)$ & $0$  & $-,+,-$ & $+,-,+$ & 0 & 0 & $\frac23$ & 0 & $-\frac13$ & $1$ & \\
\hline
\jvs  $ (1;{\bf \bar{3}},1)$ & $ (1;{\bf \bar{3}},1)$ & $0$  & $-,+,-$ & $+,-,+$ & 0 & 0 & $\frac23$ & 0 & $-\frac13$ & $-1$ & \\
\hline
\jvs   $ (1;1,{\bf 8})$ & $ (1;1,{\bf 8})$ & $0$  & $-,+,-$ & $+,-,+$ & 0 & 0 & $\frac23$ & 0 & $-\frac13$ & 0 & \\
\hline
\hline
\jvs   $ ({\bf 6};1,1)$ & $ ({\bf 5};1,1)$ & $1$  & $+,-,+$ & $-,+,-$ & $-1$ & 0 & $-\frac13$ & $-1$ & $-\frac13$ & $-1$ & \\
\jvs 		     & $ (1;1,1)$       & $1$  & $-,+,-$ & $+,-,+$ & $5$ & 0 & $-\frac13$ & $-1$ & $-\frac13$ & $-1$ &  $X_1$\\
\hline
\jvs  $ ({\bf \bar{6}};1,1)$ & $ ({\bf \bar{5}};1,1)$ & $1$  & $+,-,+$ & $-,+,-$ & $1$ & $\frac12$ & $\frac16$ & 0 & $-\frac13$ & $-1$ & \\
\jvs 		     & $ (1;1,1)$       & $1$  & $-,+,-$ & $+,-,+$ & $-5$ & $\frac12$ & $\frac16$ & 0 & $-\frac13$ & $-1$&  $\Xb_1$\\
\hline
\jvs $ (1;1,1)$ & $ (1;1,1)$ & $1$  & $+,-,+$ & $-,+,-$ & 0 & 0 & $\frac23$ & $-2$ & $-\frac13$ & $-1$ &  $Y_1$\\
\hline
\jvs $ (1;1,1)$ & $ (1;1,1)$ & $1$  & $+,-,+$ & $-,+,-$ & 0 & $\frac12$ & $-\frac56$ & $1$ & $-\frac13$ & $-1$ &  $\Yb_1$\\
\hline
\hline
\jvs $ (1;{\bf 3},1)$ & $ (1;{\bf 3},1)$ & $1$  & $-,+,-$ & $+,-,+$ & 0 & $-\frac12$ & $\frac16$ & 1 & $\frac23$ & 0 & \\
\hline
\jvs $ (1;{\bf \bar{3}},1)$ & $ (1;{\bf \bar{3}},1)$ & $1$  & $-,+,-$ & $+,-,+$ & 0 & $-\frac12$ & $\frac16$ & 1 & $-\frac13$ & 1 & \\
\hline
\jvs   $ (1;1,{\bf 8_s})$ & $ (1;1,{\bf 8_s})$ & $1$  & $+,-,+$ & $-,+,-$ & 0 & $-\frac12$ & $\frac16$ & 1 & $\frac16$ & $\frac12$ & \\
\hline
\hline
\jvs  $ ({\bf 6};1,1)$ & $ ({\bf 5};1,1)$ & $2$  & $-,+,-$ & $+,-,+$ & $-1$ & $\frac12$ & $\frac16$ & 0 & $-\frac13$ & $1$ &  \\
\jvs 	     & $ (1;1,1)$       & $2$  & $+,-,+$ & $-,+,-$ & $5$ & $\frac12$ & $\frac16$ & 0 & $-\frac13$ & $1$ &  $X_2$\\
\hline
\jvs  $ ({\bf \bar{6}};1,1)$ & $ ({\bf \bar{5}};1,1)$ & $2$  & $+,-,+$ & $-,+,-$ & $1$ & 0 & $-\frac13$ & 1 & $-\frac13$ & $1$ & \\
\jvs  		     & $ (1;1,1)$       & $2$  & $-,+,-$ & $+,-,+$ & $-5$ & 0 & $-\frac13$ & 1 & $-\frac13$ & $1$&  $\Xb_2$\\
\hline
\jvs	 $ (1;1,1)$ & $ (1;1,1)$ & $2$  & $+,-,+$ & $-,+,-$ & 0 & 0 & $\frac23$ & $2$ & $-\frac13$ & $1$ &  $Y_2$\\
\hline
\jvs	 $ (1;1,1)$ & $ (1;1,1)$ & $2$  & $-,+,-$ & $+,-,+$ & 0 & $\frac12$ & $-\frac56$ & $-1$ & $-\frac13$ & $1$ &  $\Yb_2$\\
\hline
\hline
\jvs  $ (1;{\bf 3},1)$ & $ (1;{\bf 3},1)$ & $2$  & $+,-,+$ & $-,+,-$ & 0 & $-\frac12$ & $\frac16$ & $-1$ & $-\frac13$ & $-1$ & \\
\hline
\jvs $ (1;{\bf \bar{3}},1)$ & $ (1;{\bf \bar{3}},1)$ & $2$  & $+,-,+$ & $-,+,-$ & 0 & $-\frac12$ & $\frac16$ & $-1$ & $\frac23$ & 0 & \\
\hline
\jvs  $ (1;1,{\bf 8_c})$ & $ (1;1,{\bf 8_c})$ & $2$  & $-,+,-$ & $+,-,+$ & 0 & $-\frac12$ & $\frac16$ & $-1$ & $\frac16$ & $-\frac12$ & \\
\hline
\caption[]{\rule[14pt]{1pt}{0pt}Local decomposition of ground states from the 
$T_2/T_4$ sector at $n_2=0$.
The three parities for chiral hypermultiplet components $H_L$, $H_R$ correspond to to $q_\gamma=0,\frac12,1$. 
\label{tab:states_T2T4_nn0}}
\end{longtable}
\end{center}

\begin{center}
\small
\begin{longtable}[c]{|c||c|c||c|c||c|c|c|c|c|c|c|c||c|}

\hline
\multicolumn{1}{|c||}{Bulk} &
\multicolumn{1}{c|}{$n_2=1$} &
\multicolumn{1}{c||}{ $n_3$} &
\multicolumn{1}{c|}{$H_L$} &
\multicolumn{1}{c||}{$H_R$} &
\multicolumn{1}{c|}{$t_6^1$} &
\multicolumn{1}{c|}{$t_7$} &
\multicolumn{1}{c|}{$t_8$} &
\multicolumn{1}{c|}{$t_1$} &
\multicolumn{1}{c|}{$t_2$} &
\multicolumn{1}{c|}{$t_3$} &
\multicolumn{1}{c|}{$t_4$} &
\multicolumn{1}{c||}{$t_5$\jvs} &
\multicolumn{1}{c|}{}
\\ \hline \hline
\endfirsthead

\multicolumn{14}{c}%
{{ \tablename\ \thetable{} -- continued from previous page}}  \\
\hline \multicolumn{1}{|c||}{ Bulk} &
\multicolumn{1}{c|}{$n_2=1$} &
\multicolumn{1}{c||}{ $n_3$} &
\multicolumn{1}{c|}{$H_L$} &
\multicolumn{1}{c||}{$H_R$} &
\multicolumn{1}{c|}{$t_6^1$} &
\multicolumn{1}{c|}{$t_7$} &
\multicolumn{1}{c|}{$t_8$} &
\multicolumn{1}{c|}{$t_1$} &
\multicolumn{1}{c|}{$t_2$} &
\multicolumn{1}{c|}{$t_3$} &
\multicolumn{1}{c|}{$t_4$} &
\multicolumn{1}{c||}{$t_5$\jvs} &
\multicolumn{1}{c|}{ }
\\ \hline \hline
\endhead

\endlastfoot

\multicolumn{14}{c}%
{{ \tablename\ \thetable{} -- continued on next page}} \\
\endfoot

\jvs  $ ({\bf 6};1,1)$ & $ (1,{\bf 4};1,1)$ & $0$  & $-,+,-$ & $+,-,+$ & 5 & 0 & 0 & 0 & $-\frac13$ & $1$ & $\frac23$ & 0 & \\
\jvs  			      & $ ({\bf 2},1;1,1)$ & $0$  & $+,-,+$ & $-,+,-$ & $-10$ & 0 & 0 & 0 & $-\frac13$ & $1$ & $\frac23$ & 0 & \\
\hline
\jvs  $ ({\bf \bar{6}};1,1)$ & $ (1,{\bf \bar{4}};1,1)$ & $0$  & $-,+,-$ & $+,-,+$
& $-5$ & 0 & 0 & 0 & $-\frac13$ & $-1$ & $\frac23$ & 0 & \\
\jvs   & $ ({\bf 2},1;1,1)$ & $0$  & $+,-,+$ & $-,+,-$
& $10$ & 0 & 0 & 0 & $-\frac13$ & $-1$ & $\frac23$ & 0 & \\
\hline
\jvs $ (1;1,1)$  & $ (1,1;1,1)$ & $0$  & $+,-,+$ & $-,+,-$ 
& 0 & 0 & 0 & 1 & $-\frac13$ & 0 & $\frac23$ & 0 & $Y_0$\\
\hline
\jvs $ (1;1,1)$  & $ (1,1;1,1)$ & $0$  & $+,-,+$ & $-,+,-$ 
& 0 & 0 & 0 & $-1$ & $-\frac13$ & 0 & $\frac23$ & 0 & $\Yb_0$\\
\hline
\hline
\jvs  $ (1;{\bf 3},1)$ & $ (1,1;{\bf 2},1)$ & $0$  & $+,-,+$ & $-,+,-$
& 0 & 1 & 0 & 0 & $\frac23$ & 0 & $-\frac13$ & $1$ & \\
\jvs & $ (1,1;1,1)$ & $0$  & $-,+,-$ & $+,-,+$
& 0 & $-2$ & 0 & 0 & $\frac23$ & 0 & $-\frac13$ & $1$ & $Z_0$\\
\hline
\jvs  $ (1;{\bf \bar{3}},1)$ & $ (1,1;{\bf 2},1)$ & $0$  & $-,+,-$ & $+,-,+$
& 0 & $-1$ & 0 & 0 & $\frac23$ & 0 & $-\frac13$ & $-1$ & \\
\jvs & $ (1,1;1,1)$ & $0$  & $+,-,+$ & $-,+,-$
& 0 & $2$ & 0 & 0 & $\frac23$ & 0 & $-\frac13$ & $-1$ & $\Zb_0$\\
\hline
\jvs  $ (1;1,{\bf 8})$ & $ (1,1;1,{\bf 4})$ & $0$  & $+,-,+$ & $-,+,-$
& 0 & 0 &$-1$ & 0 & $\frac23$ & 0 & $-\frac13$ & 0 & \\
\jvs   & $ (1,1;1,{\bf \bar{4}})$ & $0$  & $-,+,-$ & $+,-,+$
& 0 & 0 &$1$ & 0 & $\frac23$ & 0 & $-\frac13$ & 0 & \\
\hline
\hline
\jvs  $ ({\bf 6};1,1)$ & $ (1,{\bf 4};1,1)$ & $1$  & $-,+,-$ & $+,-,+$
& 5 & 0 & 0 & 0 & $-\frac13$ & $-1$ & $-\frac13$ & $-1$ & \\
\jvs   & $ ({\bf 2},1;1,1)$ & $1$  & $+,-,+$ & $-,+,-$
& $-10$ & 0 & 0 & 0 & $-\frac13$ & $-1$ & $-\frac13$ & $-1$ & \\
\hline
\jvs  $ ({\bf \bar{6}};1,1)$ & $ (1,{\bf \bar{4}};1,1)$ & $1$  & $-,+,-$ & $+,-,+$
& $-5$ & 0 & 0 & $\frac12$ & $\frac16$ & 0 & $-\frac13$ & $-1$ & \\
\jvs   & $ ({\bf 2},1;1,1)$ & $1$  & $+,-,+$ & $-,+,-$
& $10$ & 0 & 0 & $\frac12$ & $\frac16$ & 0 & $-\frac13$ & $-1$ & \\
\hline
\jvs $ (1;1,1)$  & $ (1,1;1,1)$ & $1$  & $+,-,+$ & $-,+,-$
& 0 & 0 & 0 & 0 & $\frac23$ & $-2$ & $-\frac13$ & $-1$ & $Y_1$\\
\hline
\jvs $ (1;1,1)$  & $ (1,1;1,1)$ & $1$  & $+,-,+$ & $-,+,-$
& 0 & 0 & 0 & $\frac12$ & $-\frac56$ & $1$ & $-\frac13$ & $-1$ & $\Yb_1$\\
\hline
\hline
\jvs  $ (1;{\bf 3},1)$ & $ (1,1;{\bf 2},1)$ & $1$  & $-,+,-$ & $+,-,+$
& 0 & 1 & 0 & $-\frac12$ & $\frac16$ & 1 & $\frac23$ & $0$ & \\
\jvs & $ (1,1;1,1)$ & $1$  & $+,-,+$ & $-,+,-$
& 0 & $-2$ & 0 & $-\frac12$ & $\frac16$ & 1 & $\frac23$ & $0$ & $Z_1$\\
\hline
\jvs  $ (1;{\bf \bar{3}},1)$ & $ (1,1;{\bf 2},1)$ & $1$  & $+,-,+$ & $-,+,-$
& 0 & $-1$ & 0 & $-\frac12$ & $\frac16$ & 1 & $-\frac13$ & $1$ & \\
\jvs & $ (1,1;1,1)$ & $1$  & $-,+,-$ & $+,-,+$
& 0 & $2$ & 0 & $-\frac12$ & $\frac16$ & 1 & $-\frac13$ & $1$ & $\Zb_1$\\
\hline
\jvs  $ (1;1,{\bf 8_s})$ & $ (1,1;1,{\bf 4})$ & $1$  & $+,-,+$ & $-,+,-$
& 0 & 0 &$1$ & $-\frac12$ & $\frac16$ & 1 & $\frac16$ & $\frac12$ & \\
\jvs   & $ (1,1;1,{\bf \bar{4}})$ & $1$  & $-,+,-$ & $+,-,+$
& 0 & 0 &$-1$ & $-\frac12$ & $\frac16$ & 1 & $\frac16$ & $\frac12$ & \\
\hline
\hline
\jvs  $ ({\bf 6};1,1)$ & $ (1,{\bf 4};1,1)$ & $2$  & $+,-,+$ & $-,+,-$
& 5 & 0 & 0 & $\frac12$ & $\frac16$ & 0 & $-\frac13$ & $1$ & \\
\jvs   & $ ({\bf 2},1;1,1)$ & $2$  & $-,+,-$ & $+,-,+$
& $-10$ & 0 & 0 & $\frac12$ & $\frac16$ & 0 & $-\frac13$ & $1$ & \\
\hline
\jvs  $ ({\bf \bar{6}};1,1)$ & $ (1,{\bf \bar{4}};1,1)$ & $2$  & $+,-,+$ & $-,+,-$
& $-5$ & 0 & 0 & 0 & $-\frac13$ & 1 & $-\frac13$ & $1$ & \\
\jvs   & $ ({\bf 2},1;1,1)$ & $2$  & $-,+,-$ & $+,-,+$
& $10$ & 0 & 0 & 0 & $-\frac13$ & 1 & $-\frac13$ & $1$ & \\
\hline
\jvs $ (1;1,1)$  & $ (1,1;1,1)$ & $2$  & $-,+,-$ & $+,-,+$
& 0 & 0 & 0 & 0 & $\frac23$ & 2 & $-\frac13$ & $1$ & $Y_2$\\
\hline
\jvs $ (1;1,1)$  & $ (1,1;1,1)$ & $2$  & $-,+,-$ & $+,-,+$
& 0 & 0 & 0 & $\frac12$ & $-\frac56$ & $-1$ & $-\frac13$ & $1$ & $\Yb_2$\\
\hline
\hline
\jvs  $ (1;{\bf 3},1)$ & $ (1,1;{\bf 2},1)$ & $2$  & $-,+,-$ & $+,-,+$
& 0 & 1 & 0 & $-\frac12$ & $\frac16$ & $-1$ & $-\frac13$ & $-1$ & \\
\jvs & $ (1,1;1,1)$ & $2$  & $+,-,+$ & $-,+,-$
& 0 & $-2$ & 0 & $-\frac12$ & $\frac16$ & $-1$ & $-\frac13$ & $-1$ & $Z_2$\\
\hline
\jvs  $ (1;{\bf \bar{3}},1)$ & $ (1,1;{\bf 2},1)$ & $2$  & $-,+,-$ & $+,-,+$
& 0 & $-1$ & 0 & $-\frac12$ & $\frac16$ & $-1$ & $\frac23$ & 0 & \\
\jvs & $ (1,1;1,1)$ & $2$  & $+,-,+$ & $-,+,-$
& 0 & $2$ & 0 & $-\frac12$ & $\frac16$ & $-1$ & $\frac23$ & 0 & $\Zb_2$\\
\hline
\jvs  $ (1;1,{\bf 8_c})$ & $ (1,1;1,{\bf 6})$ & $2$  & $-,+,-$ & $+,-,+$
& 0 & 0 & 0 & $-\frac12$ & $\frac16$ & $-1$ & $\frac16$ & $-\frac12$ & \\
\jvs & $ (1,1;1,1)$ & $2$  & $+,-,+$ & $-,+,-$
& 0 & 0 & 2 & $-\frac12$ & $\frac16$ & $-1$ & $\frac16$ & $-\frac12$ & $Z_2'$\\
\jvs & $ (1,1;1,1)$ & $2$  & $+,-,+$ & $-,+,-$
& 0 & 0 & $-2$ & $-\frac12$ & $\frac16$ & $-1$ & $\frac16$ & $-\frac12$ & $\Zb_2'$\\
\hline
\caption[]{\rule[14pt]{1pt}{0pt}Local decomposition of states from the $T_2/T_4$ sector at $n_2=1$.
The three parities for chiral hypermultiplet components $H_L$, $H_R$ correspond to
to $q_\gamma=0,\frac12,1$.}
\label{tab:states_T2T4_nn1}
\end{longtable}
\end{center}

  \clearpage
  
\begin{table}[h!]
\begin{center}
\small
\begin{tabular}{|c||c|c||c|c||c|c|c|c|c|c||c|}
\hline
Bulk & $n_2=0$ & $n_3$ &  $H_L$ & $H_R$ & $t_6^0$ & $t_1$ & $t_2$ & $t_3$ & 
$t_4$ & $t_5$ & \jvs\\
\hline
\jvs   $ (1;1,1)$ & $ (1;1,1)$ & $0$  & $-,+,-$ & $+,-,+$ & 0 & 0 & $\frac23$ & 0 & $\frac23$ & 0 & $Y_0^*$\\
\jvs         $ (1;1,1)$ & $ (1;1,1)$ & $0$  & $+,-,+$ & $-,+,-$ & 0 & 0 & $\frac23$ & 0 & $\frac23$ & 0 & $Y_0^{'*}$\\
\hline
\jvs         $ (1;1,1)$ & $ (1;1,1)$ & $1$  & $-,+,-$ & $+,-,+$ & 0 & $-\frac12$ & $\frac16$ & 1 & $-\frac13$ & $-1$& $Y_1^*$\\
\jvs         $ (1;1,1)$ & $ (1;1,1)$ & $1$  & $+,-,+$ & $-,+,-$ & 0 & $-\frac12$ & $\frac16$ & 1 & $-\frac13$ & $-1$& $Y_1^{'*}$\\
\hline
\jvs         $ (1;1,1)$ & $ (1;1,1)$ & $2$  & $+,-,+$ & $-,+,-$ & 0 & $-\frac12$ & $\frac16$ & -1 & $-\frac13$ & $1$& $Y_2^*$\\
\jvs         $ (1;1,1)$ & $ (1;1,1)$ & $2$  & $-,+,-$ & $+,-,+$ & 0 & $-\frac12$ & $\frac16$ & -1 & $-\frac13$ & $1$& $Y_2^{'*}$\\
\hline
\end{tabular}
\caption{Local decomposition of excited states from the $T_2/T_4^*$ sector at \mbox{$n_2=0$}.
The three parities for chiral hypermultiplet components $H_L$, $H_R$ correspond to
to \mbox{$q_\gamma=0,\frac12,1$}. The singlets $Y_{n_3}^*$ have oscillator numbers
$\widetilde{N}=(0,1,0)$, the $Y_{n_3}^{'*}$ have \mbox{$\widetilde{N}^* =
  (1,0,0)$}.\label{tab:states_T2T4star_nn0}} 
\end{center}
\end{table}

\begin{table}[h!]
\begin{center}
\small
\begin{tabular}{|c||c|c||c|c||c|c|c|c|c|c|c|c||c|}
\hline
Bulk & $n_2=1$ & $n_3$  & $H_L$ & $H_R$ & $t_6^1$ & $t_7$ & $t_8$ & $t_1$ & $t_2$ & $t_3$ & $t_4$ & $t_5$& \jvs\\
\hline
\jvs   $(1;1,1)$ & $(1;1,1)$ & $0$  & $+,-,+$ & $-,+,-$ & 0 & 0 & 0 & 0 & $\frac23$ & 0 & $\frac23$ & 0 
& $Y_0^*$\\
\jvs          $(1;1,1)$ & $(1;1,1)$ & $0$  & $-,+,-$ & $+,-,+$ & 0 & 0 & 0 &  0 & $\frac23$ & 0 & $\frac23$ & 0
& $Y_0^{'*}$\\
\hline
\jvs          $(1;1,1)$ & $(1;1,1)$ & $1$  & $+,-,+$ & $-,+,-$ & 0 & 0 & 0 &  $-\frac12$ & $\frac16$
& 1 & $-\frac13$ & $-1$ & $Y_1^*$\\ 
\jvs          $(1;1,1)$ & $(1;1,1)$ & $1$  & $-,+,-$ & $+,-,+$ & 0 & 0 & 0 &  $-\frac12$ & $\frac16$
& 1 & $-\frac13$ & $-1$ & $Y_1^{'*}$\\
\hline
\jvs          $(1;1,1)$ & $(1;1,1)$ & $2$  & $-,+,-$ & $+,-,+$ & 0 & 0 & 0 &  $-\frac12$ & $\frac16$
& -1 & $-\frac13$ & $1$ & $Y_2^*$\\ 
\jvs          $(1;1,1)$ & $(1;1,1)$ & $2$  & $+,-,+$ & $-,+,-$ & 0 & 0 & 0 &  $-\frac12$ & $\frac16$
& -1 & $-\frac13$ & $1$ & $Y_2^{'*}$\\ 
\hline
\end{tabular}
\caption{
Local decomposition of excited states from the $T_2/T_4^*$ sector at \mbox{$n_2=1$}.
The three parities for chiral hypermultiplet components $H_L$, $H_R$ correspond to
to \mbox{$q_\gamma=0,\frac12,1$}. The singlets $Y_{n_3}^*$ have oscillator numbers
$\widetilde{N}=(0,1,0)$, the  $Y_{n_3}^{'*}$ have \mbox{$\widetilde{N}^* =
  (1,0,0)$}.
\label{tab:states_T2T4star_nn1}}
\end{center}
\end{table}

  \clearpage
  \begin{center}
\small
\begin{longtable}{|c||c|c|c||c|c|c|c|c|c||c|}
\hline
\multicolumn{1}{|c||}{Sector} &
\multicolumn{1}{c|}{$n_2=0$} &
\multicolumn{1}{c|}{ $n_3$} &
\multicolumn{1}{c||}{ $q_\gamma$} &
\multicolumn{1}{c|}{$t_6^0$} &
\multicolumn{1}{c|}{$t_1$} &
\multicolumn{1}{c|}{$t_2$} &
\multicolumn{1}{c|}{$t_3$} &
\multicolumn{1}{c|}{$t_4$} &
\multicolumn{1}{c||}{$t_5$\jvs} &
\multicolumn{1}{c|}{}
\\ \hline \hline
\endfirsthead

\multicolumn{10}{c}%
{{ \tablename\ \thetable{} -- continued from previous page}} \\
\hline \multicolumn{1}{|c||}{Sector} &
\multicolumn{1}{c|}{$n_2=0$} &
\multicolumn{1}{c|}{ $n_3$} &
\multicolumn{1}{c||}{ $q_\gamma$} &
\multicolumn{1}{c|}{$t_6^0$} &
\multicolumn{1}{c|}{$t_1$} &
\multicolumn{1}{c|}{$t_2$} &
\multicolumn{1}{c|}{$t_3$} &
\multicolumn{1}{c|}{$t_4$} &
\multicolumn{1}{c||}{$t_5$\jvs} &
\multicolumn{1}{c|}{}
\\ \hline \hline
\endhead

\endlastfoot

\multicolumn{10}{c}%
{{ \tablename\ \thetable{} -- continued on next page}} \\
\endfoot

\jvs 	$T_1$/$T_5$   & $({\bf 10};1,1)$ & $0$ & $*$ & $\frac12$ & 0 & $-\frac16$ & $-\frac12$ & $\frac13$ & 0 & \\
\jvs 	        & $({\bf \bar{5}};1,1)$ & $0$ & $*$ & $-\frac32$ & 0 & $-\frac16$ & $\frac32$ & $\frac13$ & 0 & \\
\jvs 	        & $(1;1,1)$ & $0$ & $*$ & $\frac52$ & 0 & $-\frac16$ & $-\frac52$ & $\frac13$ & 0 & \\
\hline
\hline
\jvs   $T_1$/$T_5$    & $(1;1,{\bf 8_c})$ & $1$ & $*$ &  $\frac52$ & 0 & $-\frac16$ & $-\frac12$ & $-\frac16$ & $-\frac12$ & \\
\jvs        & $(1;{\bf 3},1)$ & $2$ & $*$ &  $\frac52$ & 0 & $-\frac16$ & $\frac32$ & $\frac13$ & 0 & \\
\jvs        & $(1;1,1)$ & $2$ & $*$ & $\frac52$ & 0 & $-\frac16$ & $\frac32$ & $-\frac23$ & $-1$ & $S_8$\\
\hline
\hline
\jvs  $T_1$/$T_5$*        & $(1;1,1)$ & $0$ & $*$ & $\frac52$ & $-\frac12$ & $-\frac23$ & $\frac12$ & $\frac13$ & 0 & $S_1$\\
\jvs 	        & $(1;1,1)$ & $0$ & $*$ & $-\frac52$ & $\frac12$ & $-\frac23$ & $-\frac12$ & $\frac13$ & 0 & $S_2$\\
\jvs 	        & $2 \times (1;1,1)$ & $0$ & $*$ &  $\frac52$ & $\frac12$ & $\frac13$ & $\frac12$ & $\frac13$ & 0 & $S_{3,4}$\\
\jvs 	        & $2 \times (1;1,1)$ & $0$ & $*$ &  $-\frac52$ & $-\frac12$ & $\frac13$ & $-\frac12$ & $\frac13$ & 0 & $S_{5,6}$\\
\jvs   & $(1;{\bf \bar{3}},1)$ & $1$ & $*$ &  $\frac52$ & 0 & $-\frac16$ & $-\frac12$ & $\frac13$ & 0 & \\
\jvs 	        & $(1;1,1)$ & $1$ & $*$ & $\frac52$ & 0 & $-\frac16$ & $-\frac12$ & $-\frac23$ & 1 & $S_7$\\
\hline
\hline
\jvs $T_3$       & $(1;{\bf 3},1)$ & $*$ & $-\frac13$ & $\frac52$ & $-\frac12$ & 0 & $\frac12$ & 0 & 1 & \\
\jvs        & $(1;{\bf \bar{3}},1)$  & $*$ & $-\frac13$ & $-\frac52$ & $\frac12$ & 0 & $-\frac12$ & 0 & $-1$  & \\
\hline
\caption[]{\rule[14pt]{0pt}{1pt}Local states from the sectors $T_1$/$T_5$ and $T_3$ at $n_2=0$. $T_1$/$T_5$* denotes oscillator
states. $S_1,S_2,S_7$ and $(1;{\bf \bar 3},1)$ from that sector
have oscillator numbers \mbox{$\widetilde{N}^* = (1,0,0)$}, $S_3$ and $S_5$ have
$\widetilde{N}^* = (0,1,0)$, $S_4$ and $S_6$ have $\widetilde{N}^* =
(2,0,0)$. \label{tab:states_T1T3T5_nn0}} 
\end{longtable}
\end{center}

\begin{center}
\small
\begin{longtable}{|c||c|c|c||c|c|c|c|c|c|c|c|c|c||c|c}

\hline
\multicolumn{1}{|c||}{Sector} &
\multicolumn{1}{c|}{$n_2=1$} &
\multicolumn{1}{c|}{ $n_3$} &
\multicolumn{1}{c||}{$q_\gamma$} &
\multicolumn{1}{c|}{$t_6^1$} &
\multicolumn{1}{c|}{$t_7$} &
\multicolumn{1}{c|}{$t_8$} &
\multicolumn{1}{c|}{$t_1$} &
\multicolumn{1}{c|}{$t_2$} &
\multicolumn{1}{c|}{$t_3$} &
\multicolumn{1}{c|}{$t_4$} &
\multicolumn{1}{c||}{$t_5$\jvs} &
\multicolumn{1}{c|}{}
\\ \hline \hline
\endfirsthead

\multicolumn{13}{c}%
{{ \tablename\ \thetable{} -- continued from previous page}} \\
\hline \multicolumn{1}{|c||}{Sector} &
\multicolumn{1}{c|}{$n_2=1$} &
\multicolumn{1}{c|}{ $n_3$} &
\multicolumn{1}{c||}{$q_\gamma$} &
\multicolumn{1}{c|}{$t_6^1$} &
\multicolumn{1}{c|}{$t_7$} &
\multicolumn{1}{c|}{$t_8$} &
\multicolumn{1}{c|}{$t_1$} &
\multicolumn{1}{c|}{$t_2$} &
\multicolumn{1}{c|}{$t_3$} &
\multicolumn{1}{c|}{$t_4$} &
\multicolumn{1}{c||}{$t_5$\jvs} &
\multicolumn{1}{c|}{}
 \\ \hline \hline
\endhead

\endlastfoot

\multicolumn{13}{c}%
{{ \tablename\ \thetable{} -- continued on next page}} \\
\endfoot

\jvs 	$T_1$/$T_5$   & $({\bf 2},1;1,1)$ & $0$ & $*$  & 0 & 1 & $-1$ & $-\frac12$ & $-\frac16$ & 0 &
$-\frac{5}{12}$ & $\frac14$ & $M_1$\\*
\jvs 		& $(1,1;1,1)$ & $0$ & $*$  & 10 & $1$ & $-1$ & $\frac12$ & $-\frac16$ & $-1$ &
$-\frac{5}{12}$ & $\frac14$ & $S_1^-$\\*
\jvs 		& $(1,1;1,1)$ & $0$ & $*$  & $-10$ & $1$ & $-1$ & $\frac12$ & $-\frac16$ & $1$ &
$-\frac{5}{12}$ & $\frac14$ & $S_1^+$\\*
\jvs 		& $({\bf 2},1;1,1)$ & $1$ & $*$  & 0 & $-1$ & $1$ & 0 & $\frac13$ & $-1$ & $\frac{1}{12}$ &
$\frac34$ & $M_2$\\*
\jvs 		& $(1,1;1,1)$ & $1$ & $*$  & $10$ & $-1$ & $1$ & $\frac12$ & $-\frac16$ & 1 & $\frac{1}{12}$
& $\frac34$ & $S_2^-$\\* 
\jvs 		& $(1,1;1,1)$ & $1$ & $*$  & $-10$ & $-1$ & $1$ & 0 & $-\frac23$ & 0 & $\frac{1}{12}$ & $\frac34$ & $S_2^+$\\*
\jvs 		& $({\bf 2},1;1,1)$ & $2$ & $*$  & 0 & $-1$ & $-1$ & 0 & $\frac13$ & $1$ & $-\frac{5}{12}$ & $\frac14$ & $M_3$\\*
\jvs 		& $({\bf 2},1;1,1)$ & $2$ & $*$  & 0 & $1$ & $1$ & 0 & $\frac13$ & $1$ & $\frac{1}{12}$ & $\frac34$ & $M_4$\\*
\jvs 		& $(1,1;1,1)$ & $2$ & $*$  & $10$ & $-1$ & $-1$ & 0 & $-\frac23$ & 0 & $-\frac{5}{12}$ & $\frac14$ & $S_3^-$\\*
\jvs 		& $(1,1;1,1)$ & $2$ & $*$  & $-10$ & $-1$ & $-1$ & $\frac12$ & $-\frac16$ & $-1$ & $-\frac{5}{12}$ & $\frac14$ & $S_3^+$\\*
\jvs 		& $(1,1;1,1)$ & $2$ & $*$  & $10$ & $1$ & $1$ & 0 & $-\frac23$ & 0 & $\frac{1}{12}$ & $\frac34$ & $S_4^-$\\*
\jvs 		& $(1,1;1,1)$ & $2$ & $*$  & $-10$ & $1$ & $1$ & $\frac12$ & $-\frac16$ & $-1$ & $\frac{1}{12}$ & $\frac34$ & $S_4^+$\\*
\hline
\hline
\jvs 	$T_3$	& $(1,1;1,1)$ & $*$ & $0$  & $10$ & $1$ & $-1$ & 0 & 0 & 2 & $\frac14$ & $\frac14$ & $S_5^-$\\
\jvs 		& $(1,1;1,1)$ & $*$ & $1$  & $10$ & $1$ & $-1$ & 0 & 0 & $2$ & $\frac14$ & $\frac14$ & $S_5^{'-}$\\
\jvs 		& $(1,1;1,1)$ & $*$ & $\frac13$  & $-10$ & $-1$ & $1$ & 0 & 0 & $-2$ & $-\frac14$ & $-\frac14$ & $S_5^+$\\
\jvs 		& $(1,1;1,1)$ & $*$ & $\frac13$  & $10$ & $1$ & $-1$ & $-\frac12$ & $-\frac12$ & $-1$ & $\frac14$ & $\frac14$ & $S_6^-$\\
\jvs 	& $(1,1;1,1)$ & $*$ & $0$  & $-10$ & $-1$ & $1$ & $\frac12$ & $\frac12$ & 1 & $-\frac14$ & $-\frac14$ & $S_6^+$\\
\jvs 		& $(1,1;1,1)$ & $*$ & $1$  & $-10$ & $-1$ & $1$ & $\frac12$ & $\frac12$ & $1$ & $-\frac14$ & $-\frac14$ & $S_6^{'+}$\\
\jvs 		& $(1,1;1,1)$ & $*$ & $-\frac13$  & $10$ & $1$ & $-1$ & $\frac12$ & $\frac12$ & $-1$ & $\frac14$ & $\frac14$ & $S_7^-$\\
\jvs 		& $(1,1;1,1)$ & $*$ & $-\frac13$  & $-10$ & $-1$ & $1$ & $-\frac12$ & $-\frac12$ & $1$ & $-\frac14$ & $-\frac14$ & $S_7^+$\\
\hline
\caption[]{\rule[14pt]{0pt}{1pt}Local states from the sectors $T_1$/$T_5$ and $T_3$ at
  $n_2=1$. \label{tab:states_T1T3T5_nn1}}
\end{longtable}
\end{center}

  \clearpage


\section{Anomaly Polynomials}\label{sec:ancoeffs} 
  In Section~\ref{sec:anomalies}, we checked that the irreducible terms in the anomaly polynomial cancel. The
  remaining piece explicitly reads
  \begin{align}
  \label{eq:I8}
    \begin{split}
      \I \left(2\pi\right)^3 I_8^{\text{bulk}}&= \frac{1}{16}\left\{ \left(\tr R^2 \right)^2 
        - \frac{1}{6} \left(\tr R^2\right) \left(\sum_A m_A \tr F_A^2 +\sum_{u,v} m_{uv} F_u F_v \right)\right. \\
      &\quad\mspace{20mu} \left. +4\sum_{A,u,v} d_{A\,uv} \left( \tr F_A^2\right) F_u F_v
        +\frac23 \sum_{u,v,w,x} h_{uvwx} F_u F_v F_w F_x\right\} \, ,
    \end{split}
  \end{align}
  with coefficients
  \begin{align}
    m_A&=\sum_{\bf r} s_A^{\bf r} v_A^{\bf r} - v_A^{\text{(adj)}} \,,& m_{uv}&= \tr_6 \left( t_u t_v
    \right) = \sum_i q_u^i  q_v^i \, , \\
    d_{A\,uv}&=\sum_{\bf r} v^{\bf r}_A \sum_{k=1}^{s_A^{\bf r}} q_u^k
    q_v^k \, , &
    h_{uvwx}&= \tr_6 \left( t_u t_v t_w t_x \right) = \sum_i q_u^i
    q_v^i q_w^i
    q_x^i  \,.
  \end{align}
  All sums are over hypermultiplets only; the vector multiplets only appear in the final term of
  $m_A$. In the sums, $i$ runs over all states, ${\bf r}$ over all representations of group $G_A$
  and $k$ over all multiplets in representation ${\bf r}$.  $q^i_u$ and $q_u^k$ are the charges of
  states and multiplets under  $U(1)_u$, and $\tr_6$ denotes the trace of the $\U1$ generators,
  i.e.~the sum over the charges of all fields. The integers $s_A^{\bf r}$  are the multiplicities of
  states transforming in that representation, and $v_A^{\bf r}$ is its quadratic index. Note that
  terms $\sim \left( \tr F_A^2\right) \left( \tr F_B^2\right)$ for two different non-Abelian factors
  $A, B$ add up to zero in our model. By explicit evaluation of these definitions in the basis $\hat{t}_u
  =  t_u/\sqrt{2} |t_u|$ we find the results  
  \begin{align}
	  m_A&=6 \left(2,2,1\right) \; , &  m_{uv}&= 6 \left( \beta_{uv} +  \delta_{uv} \right) \; , 
  \end{align}
  where $\beta_{uv}$ is given  in Eq.~(\ref{eq:beta}). Furthermore
  \begin{align}
    d_{\SU6\,uv}&=d_{\SU3\,uv}=2 d_{\SO8\,uv}=\frac12 \, \beta_{uv} \;, \\
    h_{uvwx}&= \frac{3}{2|\sigma(uvwx)|} \left( \delta_{uv} \beta_{wx} + \text{perm.} \right) \; ,  
  \end{align}
  where $|\sigma(uvwx)|$ counts all possible distinct permutations of indices $u,v,w,x$ (and only
  these are included in the bracket).

  Similarly, we calculate the local anomaly polynomial at a fixed point $f$: 
  \begin{align}
    \label{eq:I6}
    i (2 \pi)^3 I_6^{f}&= - \frac{1}{48} \sum_u m_u^{f} \, F_u \tr R^2 + \frac 12 \sum_{A,u}
    d_{Au}^{f} \,  F_u  \tr F_A^2 
    + \frac 16 \sum_{uvw} h_{uvw}^{f} \, F_u  F_v F_w .
  \end{align}
  Here the coefficients  are defined as follows:
  \begin{align}
	  m_u^f &= \tr_f \left( t_u^f \right)   = \sum_i b^i \, q_u^i \; , & 
	  d_{Au}^f &= \sum_{\bf r} v_A^{\bf r} \sum_{k=1}^{s_A^{\bf r}} b^k \, q_u^k \;, \\
    h_{uvw}^f &= \tr_f \left( t_u^f t_v^f t_w^f \right)  = \sum_i  b^i \, q_u^i q_v^i q_w^i &
  \end{align}
  All sums refer to the local spectrum at fixed point $f$, evaluated on left-handed fields. The
  local trace $\tr_f$ contains an additional factor $b^i$, which is either one for localized states
  or 1/4 for states which are induced by bulk fields; the same holds for $b^k$. We
  conveniently evaluate these expressions in a basis which consists of $\hat{t}_{\rm an}^f = t_{\rm
    an}^f/\sqrt{2} |t_{\rm an}^f|$, with $t_{\rm an}^f$  from Table~\ref{tab:u1defs},
  and orthogonal generators, $\hat{t}_1^f \equiv \hat{t}_{\rm an}^f$, $\hat{t}_{\rm an}^f \cdot \hat{t}_u^f = 0 \,(u>1)$. 
  Then the only non-vanishing terms are
  \begin{align}
    \tr_0  \hat{t}_{\rm an}^0 &=  2 \sqrt{37} \;, & 
    \tr_1  \hat{t}_{\rm an}^1 &= 2 \sqrt{10}\;  
  \end{align}
  and
  \begin{align}
    d_{\SU5 \, {\rm an}}^0 &= d_{\SU3 \, {\rm an}}^0 = 2 d_{\SO8 \, {\rm an}}^0 = 2  \tr_0 \hat{t}_{\rm
      an}^0 \left( \hat{t}_u^0 \right)^2  =  \frac{2}{3} \tr_0 \left( \hat{t}_{\rm an}^0 \right)^3  = \frac{1}{12}
    \tr_0 \hat{t}_{\rm an}^0  \;, \\
    \begin{split}
      d_{\SU2 \, {\rm an}}^1 &= d_{\SU4 \, {\rm an}}^1 =  d_{\SU2' \, {\rm an}}^1 =  d_{\SU4' \,
        {\rm an}}^1 \\
      & = 2  \tr_0  \hat{t}_{\rm an}^1 \left(\hat{t}_u^1 \right)^2 = \frac{2}{3}  \tr_0 \left(
        \hat{t}_{\rm an}^1 \right)^3= 
      \frac{1}{12}  \tr_0 \hat{t}_{\rm an}^1\;.  
    \end{split}
  \end{align}

  This shows explicitly that both anomaly polynomials factorize in the required way,
  Eq.~(\ref{eq:Ireducible}), i.e.~the Green-Schwarz universality relations with levels $\alpha_{\SO N} = 1$
  and $\alpha_{\SU N} = 2$ are fulfilled,
  \begin{align}
    \frac{1}{48}   \tr_f  \hat{t}_{\rm an}^f   &= \frac16   \tr_f \left( \hat{t}_{\rm an}^f \right)^3 
    = \frac12 \tr_f \hat{t}_{\rm an}^f \left(  \hat{t}_u^f \right)^2 
    = \frac{1}{2 \, \alpha_A} d_{ A \, {\rm an} }^f \;.
  \end{align}
  

\clearpage


\addcontentsline{toc}{section}{References}


\end{document}